\documentclass[prc,twocolumn,preprintnumbers,showpacs,superscriptaddress]{revtex4}
\usepackage[colorlinks=true,linkcolor=blue,citecolor=blue,urlcolor=darkred]{hyperref}
\usepackage[T1]{fontenc}
\usepackage{amsmath}
\usepackage{natbib}
\usepackage{soul}
\usepackage{xcolor}
\usepackage{lipsum}
\usepackage{float}
\usepackage{graphicx}	
\usepackage{multirow,makecell}
\usepackage{placeins}
\usepackage{bm,amsfonts,hhline,amssymb,microtype,float,latexsym,epsf,mathtools,times,makecell,array}
\usepackage{tabularx}
\usepackage{epstopdf}
\usepackage{gensymb}
\usepackage{pifont}
\usepackage{hyperref}
\usepackage{lineno}
\usepackage{natbib}

\definecolor{ForestGreen}{HTML}{228B22}
\newcommand{\aap}{Astron.\& Astrophys.}
\newcommand{\mnras}{Mon. Not. R. A. S.}
\newcommand{\nphysa}{Nucl. Phys. A}

\definecolor{darkred}{rgb}{0.7,.1,.2}

\begin{document}
\title{Superconductivity in Magnetars: Exploring Type-I and Type-II States in Toroidal Magnetic Fields}

\author{Mayusree Das}
\email{mayusreedas@iisc.ac.in}
\affiliation{Joint Astronomy Programme, Department of Physics, Indian Institute of Science, Bangalore, 560012, India}

\author{Armen Sedrakian}
\email{armen.sedrakian@uwr.edu.pl}
\affiliation{Frankfurt Institute for Advanced Studies, D-60438 Frankfurt am Main, Germany}
\affiliation{Institute of Theoretical Physics, University of Wroclaw, 50-204 Wroclaw, Poland}

\author{Banibrata Mukhopadhyay}
\email{bm@iisc.ac.in}
\affiliation{Department of Physics, Indian Institute of Science, Bangalore, 560012, India}
\affiliation{Joint Astronomy Programme, Department of Physics, Indian Institute of Science, Bangalore, 560012, India}


\begin{abstract}  We present a first two-dimensional general-relativistic analysis of superconducting regions in axially symmetric highly magnetized neutron star (magnetar) models with toroidal magnetic fields. We investigate the topology and distribution of type-II and type-I superconducting regions for varying toroidal magnetic field strengths and stellar masses by solving the Einstein-Maxwell equations using the XNS code. Our results reveal that the outer cores of low- to intermediate-mass magnetars sustain superconductivity over larger regions compared to higher-mass stars with non-trivial distribution of type-II and type-I regions. Consistent with previous one-dimensional (1D) models, we find that regardless of the gravitational mass, the inner cores of magnetars with toroidal magnetic fields are devoid of $S$-wave proton superconductivity. Furthermore, these models contain non-superconducting, torus-shaped regions—a novel feature absent in previous 1D studies.  Finally, we speculate on the potential indirect effects of superconductivity on continuous gravitational wave emissions from millisecond pulsars, such as PSR J1843-1113, highlighting their relevance for future gravitational wave detectors.  \end{abstract}

\maketitle

\section{Introduction}
\label{sec:intro}

The onset of superconducting and superfluid states in compact stars, particularly neutron stars (NSs), occurs shortly after their formation, as they cool below the critical temperature $T_c \sim 10^9 \mathrm{~K}$, typically within hours of their birth in a supernova explosion~\cite{Page06,Page2011PhRvL,Blaschke2013PhRvC}. Finite-temperature effects, near but below the critical temperature, are significant, particularly in the context of neutrino emission from compact stars. Moreover, the proton fluid in the core of a compact star is highly degenerate, with a Fermi temperature $T_F \ge 10^{11} \mathrm{~K} \gg T_c$ in the relevant density range. This strong degeneracy condition is essential for the applicability of the microscopic theory of superconductivity~\cite{Abrikosov:Fundamentals}. It implies that the superconductor is well within the weak-coupling  regime of pairing with a coherence length much larger than the interparticle distance~\cite{Sedrakian2019EPJA}. 

The motivation to study the superfluidity and superconductivity of nucleonic fluids in compact stars arises from their influence on many astrophysical manifestations. They range from their neutrino and photon cooling~\cite{Page06} to pulsar glitches~\cite{Chamel2017,Haskell18,GlitchesZhou2022,Antonopoulou2022RPPh} to the evolution of the magnetic field(s) ~\cite{Pons2009AA,Ascenzi2024MNRAS} and the generation of gravitational waves~\cite{Alford2014ApJ,Piccinni2022,Andersson2021Univ}.

Proton superconductors can exist in either a type II state, characterized by an array of quantized fluxtubes between the lower and upper critical fields, or a type I state, where superconducting and normal regions form interpenetrating domains (for reviews and references, see Refs.~\cite{Graber2017,Haskell18}.  The field strengths of highly magnetized NSs (magnetars) $\left(\gtrsim 10^{15} \mathrm{G}\right)$ disrupt $S$ wave superconductivity and superfluidity in these objects. For neutral fluids, the field aligns the spins of the neutrons along their direction, breaking the opposite-spin pairing necessary to form Cooper pairs~\cite{Stein2016PhRvC}. In the case of protons, superconductivity is suppressed when the radius of curvature of the proton orbits, bent by the field exceeding the critical field strength, becomes smaller than the coherence length of Cooper pairs~\cite{Sinha2015PhRvC}.
Due to the coupling of the proton superconductor to the neutron superfluid,  a correction to the upper critical field arises,
which was studied via Ginzburg-Landau (GL) theory in Ref.~\cite{Sinha2015PhRvC}. A general phase diagram of type II and type I phases for coupled neutron and proton superfluids within the GL formalism {including both upper and lower critical fields} was studied in Ref.~\cite{Haber2017PhRvD}. {Here, we will neglect for simplicity the corrections arising due to coupling the proton superconductor to the neutron superfluid through the entrainment discussed in these references. }

Previous studies \cite{Sinha2015PPN, Sinha2015PhRvC} of the effects of a strong magnetic field on type-II and type-I superconducting regions in magnetars used one-dimensional models with prescribed field profiles, typically assuming a parametrization of poloidally dominated dipole field~\cite{Chatterjee2019PhRvC}.
Additionally, Ref.~\cite{Sinha2015PhRvC} showed, using again one-dimensional equilibrium models, that type-II superconductivity emerges in the outer core starting at the crust-core interface, followed by type-I superconductor deeper within the star. The inner core is typically void of $S$-wave proton superconductivity.  {Here, we will focus exclusively on proton $S$-wave superconductivity. At higher densities and for higher Fermi-energies, $P$-wave proton pairing is not excluded but is likely to be weak~\cite{Raduta2019MNRAS}. Similarly, in nearly isospin symmetric matter at high densities, neutron-proton $D$ wave pairing may emerge, but its existence requires nucleonic matter close to the zero isospin state~\cite{Alm1996NuPhA}.}

This work aims to extend these models by (a)~assuming axial symmetry of {\it stellar field $B_S$} to create two-dimensional representations of superconducting regions; and (b)~focusing on magnetars with dominant toroidal field configurations,
that are more stable than those dominated by poloidal fields (purely poloidal fields are known to be inherently unstable); (c)~we further employ temperature-dependent gap function, which makes our study applicable to finite temperatures relevant for cooling compact stars. To achieve these goals, we use the XNS code~\cite{sold2021main} to solve the Einstein-Maxwell equations, employing a realistic equation of state (EoS) for dense nucleonic matter and pairing gap profiles derived from microscopic calculations. 

{ 
We do not incorporate the back-reaction of superconductivity on the equilibrium and stability of magnetars.
At the level of the energies involved, this is justified by the fact that the energy density associated with the superconductivity is much smaller than the characteristic Fermi energy of fermions. A further point is that the superconductivity affects electromagnetism and the overall field as the induction differs from the magnetic field intensity. The effect of this difference on the global structure of magnetars is ignored here. Finally, the XNS solver is used in a single fluid approximation, i.e., the multifluid nature of neutron-proton mixtures and their mutual entrainment is neglected. Two fluid approaches were developed for low-field neutron stars to study toroidal field configurations in the Newtonian approximation~\cite{AW2008MNRAS,Lander12}.}

As a possible application of our results, we discuss the emission of gravitational waves (GWs) from millisecond pulsars (MSPs) and the potential influence of type-II superconductivity of the interiors of compact stars.

\section{Critical temperature and critical magnetic fields of superconductors}
\label{sec:Tc}
\begin{figure}[bt]
\centering
\includegraphics[width=\columnwidth]{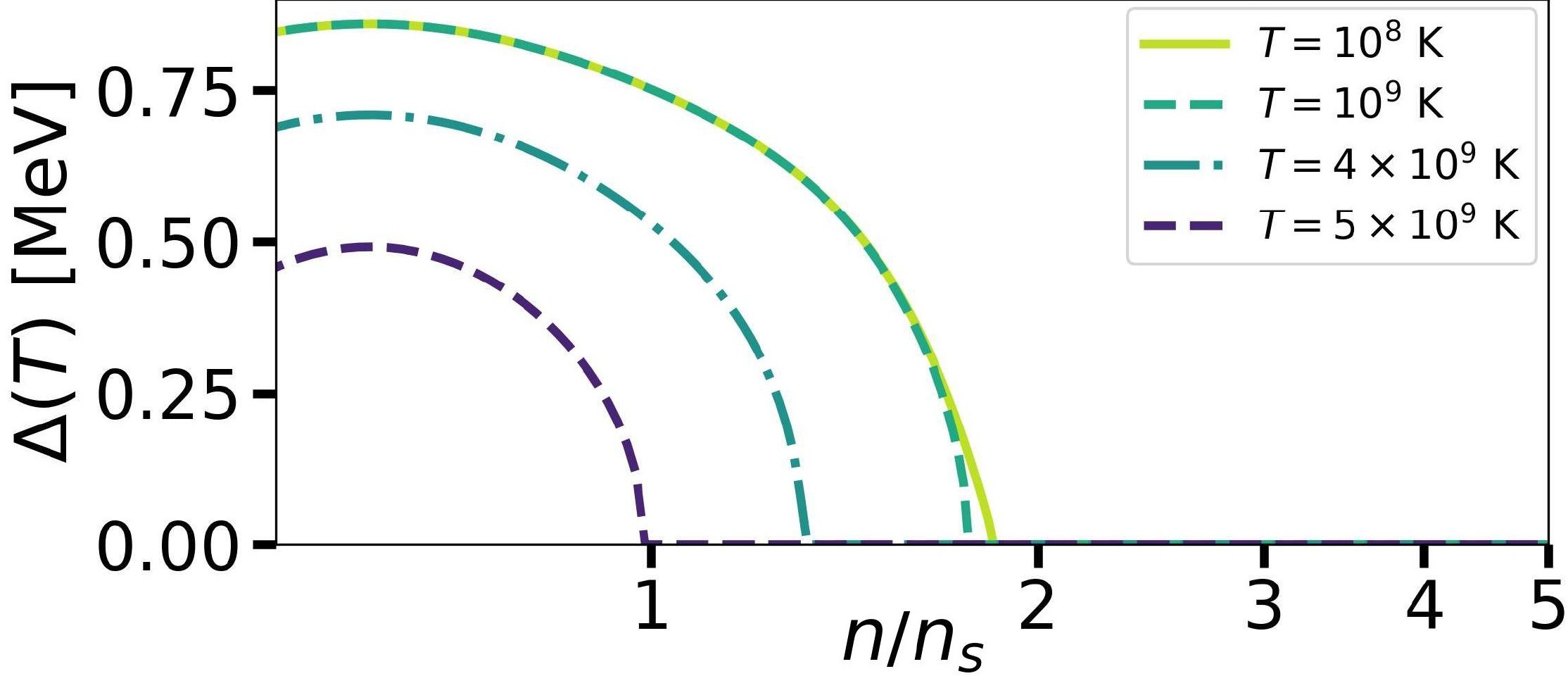}
\caption{  Dependence of proton pairing gap $\Delta$ on density at various temperatures
  indicated in the plot. }
\label{fig:1}
\end{figure}
Following previous work \cite{Sinha2015PhRvC,Muhlschlegel1959ZPhy}, the gap at a finite temperature $T$ is
\begin{equation}
  \label{eq:D_approx}
\frac{\Delta(\tau)}{\Delta(0)} = 
\begin{cases}
\displaystyle 1 - \sqrt{2\gamma \tau} e^{-\pi / (\gamma \tau)}, & 0 \leq \tau \leq 0.5, \\[0pt]
\displaystyle \sqrt{3.016(1-\tau) - 2.4(1-\tau)^2}, & 0.5 < \tau \leq 1,
\end{cases}
\end{equation}
where $\tau=T/T_c$ with $T_c$ being the critical temperature of an $S$-wave superconductor, $\gamma=1.781$, $\Delta(0)=1.76 k_B T_c$ is the pairing gap at zero temperature for an $S$-wave superconductor and $k_B$ is the Boltzmann constant.
To map these quantities onto 
the structure of a magnetar, we adopt the ``GPPVA-DDME2''~\cite {gppva,Lala2005} EoS of nucleonic matter  from {\sc CompOSE} repository~\cite{CompOSE2022EPJA,Dexheimer2022}.

Fig.~\ref{fig:1} shows the proton gap $\Delta$ as a function of density for several temperature values, calculated from Eq. \eqref{eq:D_approx}. 
It is seen that the gap, and, therefore, the superconducting area within the star, grows as it cools from $T \sim 10^{10}$~K down to $T\sim 10^{8}$~K during the neutrino cooling era which takes about $10^4$ years~\cite{Page06}.  Note that the protons are bound in clusters inside the crust of a magnetar, therefore, mesoscopic and bulk superconducting effects are not present for densities $n\le 0.5n_s$ corresponding to the crust-core boundary.

  Superconductors on mesoscopic scales are characterized by two scales - the penetration depth $\lambda$ of the magnetic field in the superconductor and the coherence length $\xi$ characterizing the size of a Cooper pair. Their
  electrodynamics can be treated phenomenologically in terms of the GL theory, which contains as a key input the 
  GL parameter $\kappa = \lambda / \xi $~\cite{Abrikosov:Fundamentals}. 
  {In the following, we use $H$ for local magnetic field intensity ($H$-field) and $B$ for local magnetic field induction, which includes the response of the superconductor to an applied field}, see Refs.~\cite{Sedrakian1995ApJ,Haber2017PhRvD}. For $\kappa > 1/\sqrt{2}$, the surface energy between the normal and superconducting states is negative and for the local field induction between the lower and upper critical fields: $H_{c1}\le H\le H_{c2}$, superconductor sustains quantized fluxtubes which carry a quantum of circulation. For $H> H_{c2}$ superconductivity breaks down.  Conversely, for $H < H_{c1}$, the field is supposed to be expelled from the superconducting regions; however, metastable fluxtube featuring state, in this case, was conjectured in Ref.~\cite{Baym1969}, as the flux expulsion timescale is rendered very long by the high electric conductivity of core material. {(If the magnetic field is assumed to pre-exist during the onset of superconductivity, it is treated as the local field intensity $H$  to which the critical fields $H_{c1}$ and $H_{c2}$ are compared.)} As the density increases $\kappa$ decreases below the limiting value $1/\sqrt{2}$ in which case type-I superconducting state is preferable. In this state, normal and superconducting states form layers of normal-superconducting domains whose size and structure depend on various factors, including nucleation dynamics, magnetic history, etc.

  \subsection{Type-II superconducting state}
  Proton superconductors may exist in a type-II state
  between the lower ($H_{c1}$) and upper ($H_{c2}$) critical fields, 
see Refs.~\cite{Baym1969,Sedrakian1995ApJ,Glampedakis2011,Gusakov2016a,Rau2020} and references therein. 
  The critical fields are given by
  \begin{equation}
    \label{eq:Hc1}
 H_{c 1}=\frac{\Phi_0}{4 \pi \lambda^2} \left[\ln \kappa +C_1 (\kappa)\right], \quad  H_{c 2}=\frac{\Phi_0}{2 \pi \xi^2} ,
\end{equation}
where {$C_1 (\kappa) = 0.5 + {(1 + \ln 2)}/{(2\kappa - \sqrt{2} + 2)}$, arises from numerical computations of the exact flux-tube
energy in a type-II superconductor} \cite{Hu1972, Brandt2003}, 
$\Phi_0=\pi\hbar c/e$ is the flux quantum, $\hbar$ is reduced Planck's constant, $c$ is speed of light, and $e$ is electron's charge.
The mesoscopic parameters entering these equations are given by
\begin{equation}
\xi=\frac{\hbar^2 k_{F_p} }{ \pi m_p^* \Delta }, \quad 
\lambda = \sqrt{\frac{m_p^{*2} c^2}{4 \pi e^2 \rho_{p}}},
\end{equation}
where $ k_{F_p}$ is the proton Fermi wave-number,  
  $\rho_p$ is the density of proton fluid, $m_p^* $  is  the proton effective mass.

  \subsection{Type-I superconductivity }

  Type-I superconducting proton fluid
  undergoes a transition to the normal state above the thermodynamic critical field $H_{\rm cm}$, given by
  \begin{equation}
    \label{eq:Hcm}
    H_{\rm cm}=  \frac{\kappa\Phi_0}{2^{3/2}\pi \lambda^2 }  =  
    \frac{\Phi_0}{2^{3/2}\pi \xi^2 \kappa} = \frac{H_{c 2}}{\sqrt{2} \kappa}.
  \end{equation}
The $H$-field is either expelled from the type-I superconductor or, instead,
  nucleates in the form of alternating domains of normal and superconducting regions, whose shape and size could depend on several factors, such as the presence of magnetized vortices in the neutron fluid and entrainment effect~(see Refs.~\cite{Sedrakian1997MNRAS,Sedrakian2005PhRvD}).
 
\section{Magnetar models with type-I and type-II superconducting regions 
}

We employ the general relativistic magneto-hydrostatic solver XNS 4.0~\cite{sold2021main} to model magnetized, axially symmetric compact stars. {
The XNS solver takes as input the central density, maximum magnetic field, magnetic field configuration (poloidal, toroidal, or mixed), and the star’s angular rotation frequency. It numerically solves the Einstein-Maxwell equations, yielding the stellar mass and magnetic field profile as outputs. While the rotation frequency is finite (we assume uniform rotation), for the present purpose we consider it negligibly small—consistent with observed magnetar candidates—such that its effect on the stellar structure is insignificant, i.e.,  the star's physical properties depend solely on the radius ($r$) and the polar angle ($\theta$).}
\begin{figure}[hbt]
  \centering
\includegraphics[width=\columnwidth]{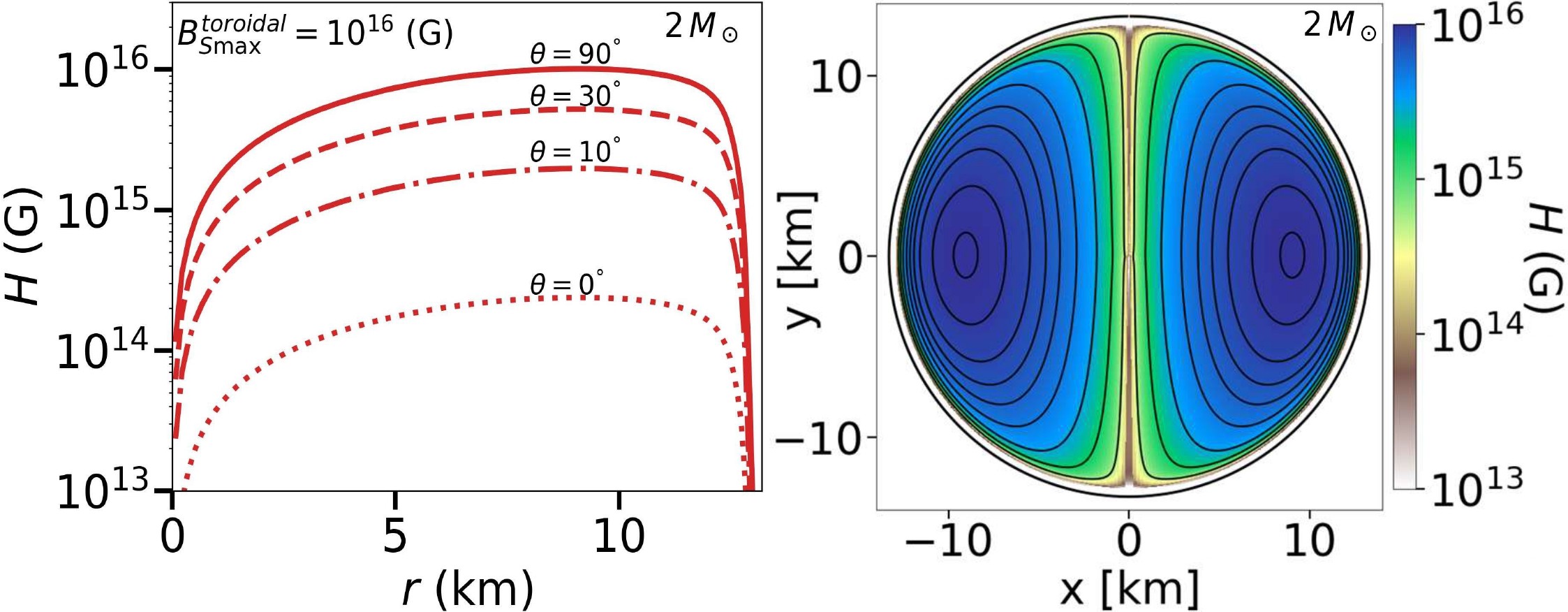}
\caption{Left panel: The magnitude of the toroidal field as a
function of radius for fixed values of the polar angle,  $\theta$, in spherical coordinates, for $M=2M_\odot$.
Right panel: A color plot illustrates the distribution of the $H$-field in the 
$x-y$ plane, with the magnetic axis aligned along the $y$-direction in Cartesian coordinates.
Solid black lines represent the isocontours of the field.}
\label{fig:2}
\end{figure}

As is well known, the stellar structure can be decoupled from the thermal evolution of the star, and finite-temperature effects can be neglected. Consequently, the structure can be computed using the adopted zero-temperature EoS ``GPPVA-DDME2''~\cite {gppva,Lala2005}. Once the stellar configuration is determined, we map the temperature-density-dependent gap onto the stellar structure.

In this work, we focus exclusively on models with toroidally dominated magnetic fields. The motivation for selecting a toroidal field configuration is as follows: It is well established that compact stars with purely toroidal $B_T$ or purely poloidal $B_P$ fields are inherently unstable. Consequently, mixed field models, where the ratio of the poloidal field energy ($E_{M}^P$) to the total field energy ($E_{M}^P + E_{M}^T$) ranges from $10^{-3}$ to $0.8$, have been identified as stable~\cite{BR2009}. This indicates that toroidally dominated fields are more probable for stable stellar models and, thus,  represent the realistic stable equilibria
to model magnetars. Fig.~\ref{fig:2} shows the radial profile of the toroidal local $H$-field for various fixed polar angles $\theta$ in spherical coordinates (left panel) and $x$--$y$ cut orthogonal to the $z$ axis in Cartesian coordinates with the $y$-axis along the magnetic axis. It is seen that the largest field strengths are achieved along the equatorial radius for $\theta=90^\circ$ and the field strength is minimal in the orthogonal direction along the magnetic axis. Note that for each fixed $\theta$, the field attains its maximum away from the star's center and within the outer core, coincidentally where the proton superconductivity is most prominent.
\begin{figure}[hbt]
\centering
    \includegraphics[width=\columnwidth]{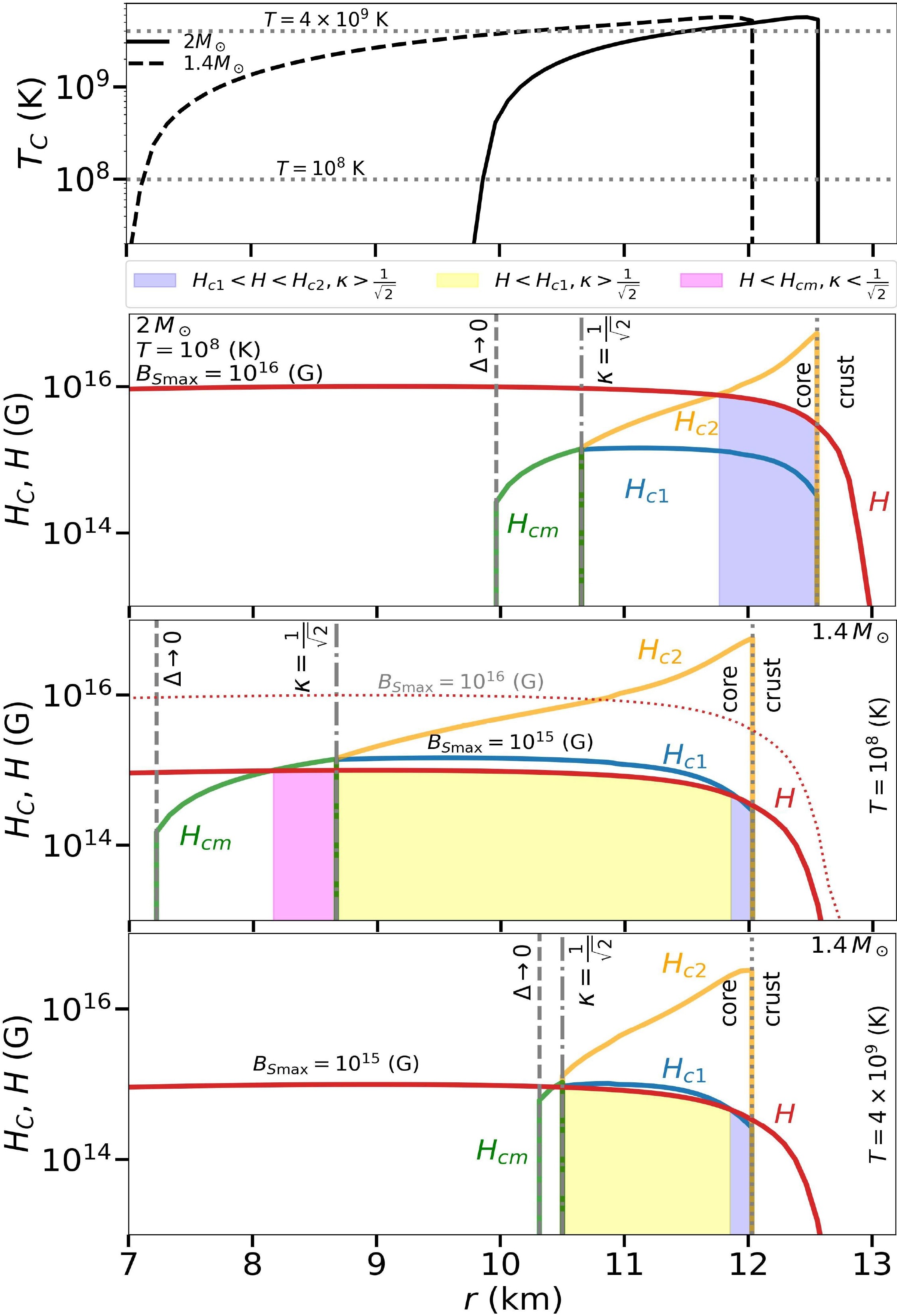}
    \caption{
      The dependence of the critical temperature and critical fields on the radial coordinate in spherical coordinates is shown for a fixed polar angle of $\theta = 90^\circ$, which corresponds to the equator radius, in models of varying mass and temperature. The top panel illustrates the critical temperature, $T_c$, for stars with masses of $1.4M_\odot$ and $2M_\odot$. The subsequent three panels depict the corresponding critical fields $H_{c1}$, $H_{c2}$, and $H_{\rm cm}$, along with
      the $H$-field, as determined by solutions to the Einstein-Maxwell equations, for indicated maximum field strength, $B_{S\text{max}}$ and temperature. Two vertical lines mark the transitions between type-I and type-II superconductivity
       at $\kappa = 1/\sqrt{2}$ and the transition from superconducting to unpaired ($\Delta \to 0$) state. The shaded regions highlight the following: For type-II superconductors, the fluxtube array state, where $H_{c1} < H < H_{c2}$ and Meissner state where $H<H_{c1}$; For type-I superconductors, the Meissner or layered-domain state, where $H < H_{cm}$.
}
\label{fig:3}
\end{figure}

\begin{figure}
\centering
  \includegraphics[scale=0.5]{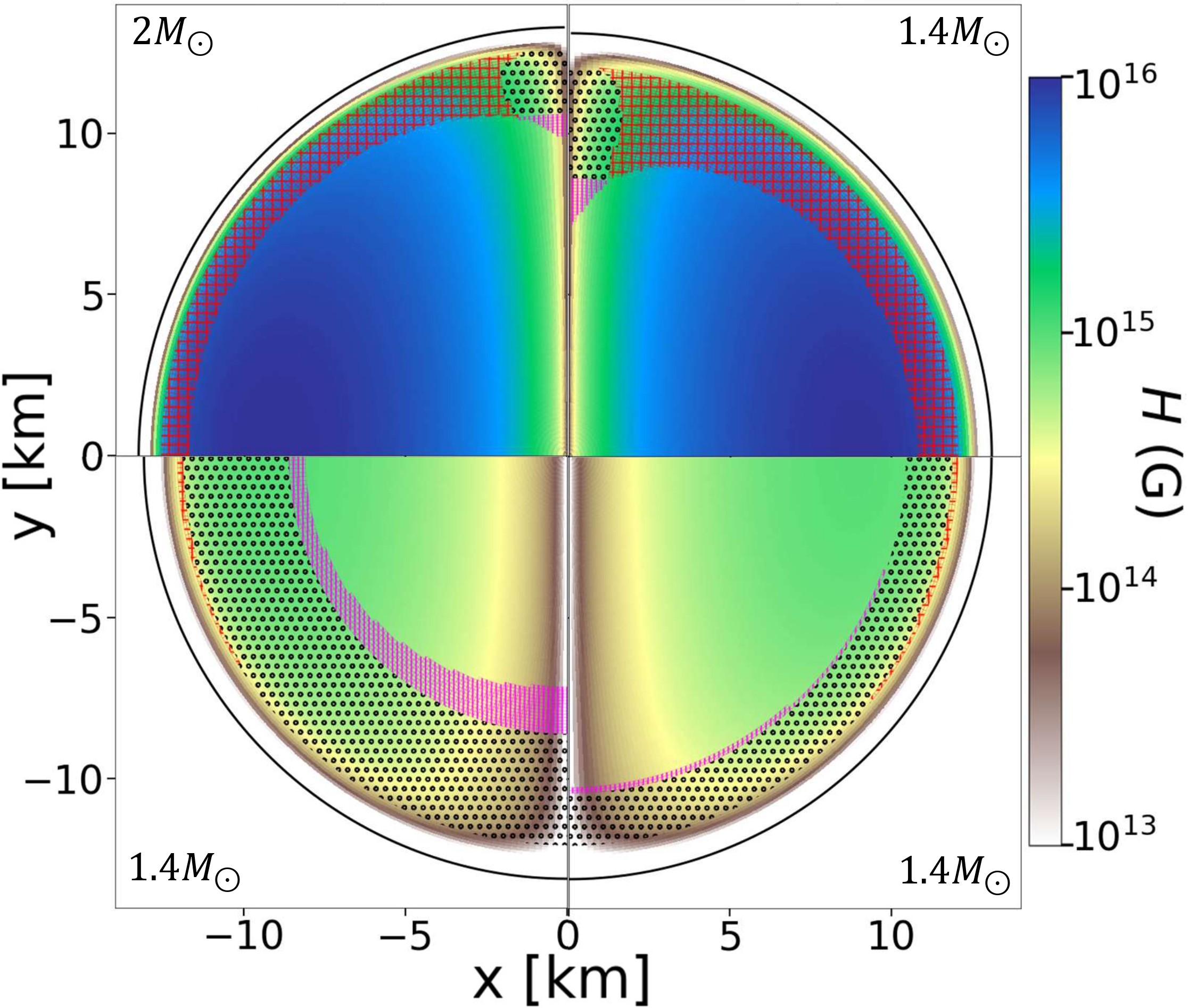}
    \caption{
Color plot illustrating the distribution of the toroidal $H$-field in the $x-y$ plane, where the magnetic axis aligns with the $y$-direction in Cartesian coordinates. The plot is divided into four quadrants, each corresponding to specific parameter values. Top-left: $M=2 M_{\odot}, B_{S \max }=10^{16} \mathrm{G}, T=10^8 \mathrm{~K}$;
top-right: $M=1.4 M_{\odot}, B_{S \max }=10^{16} \mathrm{G}, T=10^8 \mathrm{~K}$; bottom-left: $M=1.4 M_{\odot}, B_{S \max }=10^{15} \mathrm{G}, T=10^8 \mathrm{~K}$;
bottom-right: $M=1.4 M_{\odot}, B_{S \max }=10^{15} \mathrm{G}, T=4 \times 10^9 \mathrm{~K}$. The different superconducting regions are identified as follows:
Red crosses denote the type-II superconducting region, where $\kappa>1 / \sqrt{2}$ and $H_{c 1}<H< H_{c 2}$, corresponding to the presence of fluxtubes;
black dots indicate the Meissner state within the type-II superconducting region, characterized by $\kappa>1 / \sqrt{2}$ and $H<H_{c1}$;
magenta vertical hatching represents the Meissner or layered-domain state in the type-I superconducting region, where $\kappa<1 / \sqrt{2}$ and $H<H_{c 1}$.
Note that the magnetic induction $B=0$ in both type-I and type-II regions when $H$ is below the corresponding lower critical field. However, we distinguish these regions due to the potential nucleation of normal domains or fluxtubes if field expulsion timescales are long~\cite{Baym1969}.
The unhatched regions denote nonsuperconducting states, which occur when $H > H_{cm}$ in the type-I region or $H > H_{c2}$ in the type-II region.}
\label{fig:4}
\end{figure}

Fig.~\ref{fig:3} (upper panel) illustrates $T_c=\Delta(0)/1.76k_B$ calculated from the zero-temperature gap for proton pairing as a function of stellar radius for two stellar models with masses of $M=2 M_\odot$ and $1.4 M_\odot$, corresponding to central densities of $7.7 \times 10^{14}$ and $5.5 \times 10^{14}$ g cm$^{-3}$, respectively. Due to the rapid decrease of $T_c$ at high densities, the density threshold above which $S$-wave proton superconductivity vanishes is insensitive to the actual value of the temperature. Note that $T_c$  is cut off at the crust-core boundary, where protons are clustered in nuclei.  For the $2 M_\odot$ model, the superconducting region is confined to a narrow range of $r$: {$10 \lesssim r \lesssim 12.5$ km for $T=10^8$ K}. In contrast, the $1.4 M_\odot$ model exhibits a broader superconducting region, spanning $7 \lesssim r \lesssim 12$ km. This difference arises because the low-density region, which is favorable for pairing, is more extended in the $1.4 M_\odot$ case. {We then proceed assuming the temperature within the phenomenologically relevant range of $10^8 \leq T \leq 4\times10^9$ K such that $T<T_c$.} 

The extent of the superconducting region, in the adopted range of $T$, is further determined by the local $H$-field and the critical fields $H_{c1}$ and $H_{c2}$, for a type-II superconductor, and $H_{\rm cm}$ for a type-I superconductor. Specifically, the following options arise:
\begin{itemize}
\item Region exhibiting type-II superconductivity, $\kappa > 1/\sqrt{2}$:
  
     $H> H_{c2}$,  normal state is preferred;
     $H_{c1} \leq H \leq H_{c2}$,  superconductor features fluxtube arrays {with magnetic field induction $B\simeq H$ except when $H\to H_{c1}$};
   $H < H_{c1}$, the Meissner state is preferred over the mixed state {with $B=0$}, but an (inhomogenous) lattice of fluxtubes
      may arise if the expulsion time is too long {with $B\simeq H$}. 
   
\item Region exhibiting type-I superconductivity, $\kappa < 1/\sqrt{2}$:

    $H > H_{\rm cm}$,  normal state is preferred;
     $H < H_{\rm cm}$, the Meissner state {with $B=0$} is preferred which may, under specific conditions, break down into 
      domains of normal and superconducting states {with $B\neq 0$.}
   
  \end{itemize}
  Returning to Fig.~\ref{fig:3}, the lower three panels illustrate the critical field values $H_{c1}, H_{c2}$, and $H_{\rm cm}$ (using Eqs. \eqref{eq:Hc1} and \eqref{eq:Hcm}), as functions of the equatorial radial coordinate for $\theta=90^{\circ}$.  These are presented for combinations of two masses $\left(1.4 M_{\odot}\right.$ and $2 M_{\odot}$), two temperatures ($T=10^8 \mathrm{~K}$ and $T=4 \times 10^9 \mathrm{~K}$), and maximal stellar field values $B_{S \max }=10^{15} \mathrm{G}$ and $B_{S \max }=10^{16} \mathrm{G}$, to analyze the effects of varying these parameters on the type-I and type-II superconducting regions. In each panel, the shaded areas correspond to the various regimes of type-I and type-II superconductivity, as determined by the criteria outlined earlier.  Moving outward from the central regions toward the surface, two transitions occur:
(1) the normal-superconducting phase transition, where $\Delta$ becomes non-zero;
(2) the transition from type-I to type-II superconductivity at the point $\kappa=1 / \sqrt{2}$.
  The following trends are observed:
(a) Models with larger masses exhibit a narrower range of the superconducting region (compare the second and third panels of Fig. \ref{fig:3}).
(b) For very large $H$, where $H$ does not intersect with the critical fields, superconductivity is absent. If $H$ intersects only $H_{c 2}$, a type-II superconducting fluxtube state emerges near the crust-core interface (Fig. \ref{fig:3} panel 2). When $H$ intersects only $H_{c 1}$, the type-II flux tube region narrows down and shifts to larger radii, along with a type-II Meissner state forming near the core (Fig. \ref{fig:3} panel 3).
 For low $H$ that does not intersect with $H_{c 1}$, the Meissner state becomes energetically favorable. (c) Increasing the temperature while the remaining parameters are fixed shrinks the domain of the superconducting regions as the gap decreases but does not affect the arrangement of different phases. 
The analysis for the type-I superconducting region is more straightforward: as the field decreases, a Meissner or layered-domain phase emerges at the radius where $\kappa=1 / \sqrt{2}$ and extends radially inward till the region where $\Delta=0$
(see the fourth panel of Fig. \ref{fig:3}, where the domain appears as a tiny zone, hardly visible).

We now turn to the key point of this work by extending the discussion from the previous section in 1D to the case of 2D axially symmetric models. We assume that the temperature of the star is below $T_c$ for proton pairing, which is a realistic scenario a few hours after its birth when the core temperature is $T \leq T_c$.

Once the models are computed using the XNS solver for the Einstein-Maxwell equations, we identify the type-I, type-II, and non-superconducting regions according to the criteria outlined above. The results are summarized in Fig.~\ref{fig:4}, where we choose $M=1.4 M_{\odot}$ and $2 M_{\odot}$, $T=10^8 \mathrm{~K}$ and $T=4 \times 10^9 \mathrm{~K}$, and $B_{S \max }=10^{15} \mathrm{G}$ and $B_{S \max }=10^{16} \mathrm{G}$. The trends observed in the 1D representation in Fig.~\ref{fig:3} are intact along radial directions characterized by large angles. On the contrary, for angles near $\theta=0^{\circ}$, the $H$ field is significantly weaker, leading to differences in the field distribution and phase topology. For instance, in the $2 M_{\odot}$ model (upper left panel), there is a narrow type-II fluxtube region near the crust-core interface at the equator. This region becomes more extensive as one moves toward the poles. Ultimately, at the magnetic poles, a type-II Meissner state is favored, followed by a type-I state at higher densities. In comparison, the $1.4 M_{\odot}$ model (upper right panel) has larger superconducting regions with essentially the same topology and structure of superconducting regions. 
  These features arise from the interplay between the toroidal field structure and the radial variation of the critical fields, leading to the complex geometries of the superconducting regions. 
 A comparison between two $1.4 M_{\odot}$ models with different fields (upper right and lower left panel) shows that the high $H$-field results in a torus-shaped region being void of superconductivity where the field is maximal. 
 Also, the topology of superconducting regions is different: for the lower field, it has an ellipsoidal shape, whereas for the higher field model, the ellipsoid has open poles. 
 A comparison between two $1.4 M_{\odot}$ models with different temperatures (lower left and right panels) shows that the higher temperature means a narrower superconducting region, as expected.
Additionally, the superconducting regions differ, as best seen in the lower left panel, where a prolate type-I region extends most prominently at the poles.
A common feature across all panels is that the inner core remains nonsuperconducting due to the phase transition from the $S$-wave superconducting state to a nonsuperconducting state at densities $\rho>4.3 \times 10^{14} \mathrm{~g} \mathrm{~cm}^{-3}$. A star would only remain superconducting down to its center in the case of very low mass.

\section{Observational Implications}
\label{sec:msp}
\begin{figure}
\begin{center}
\includegraphics[width=\columnwidth]{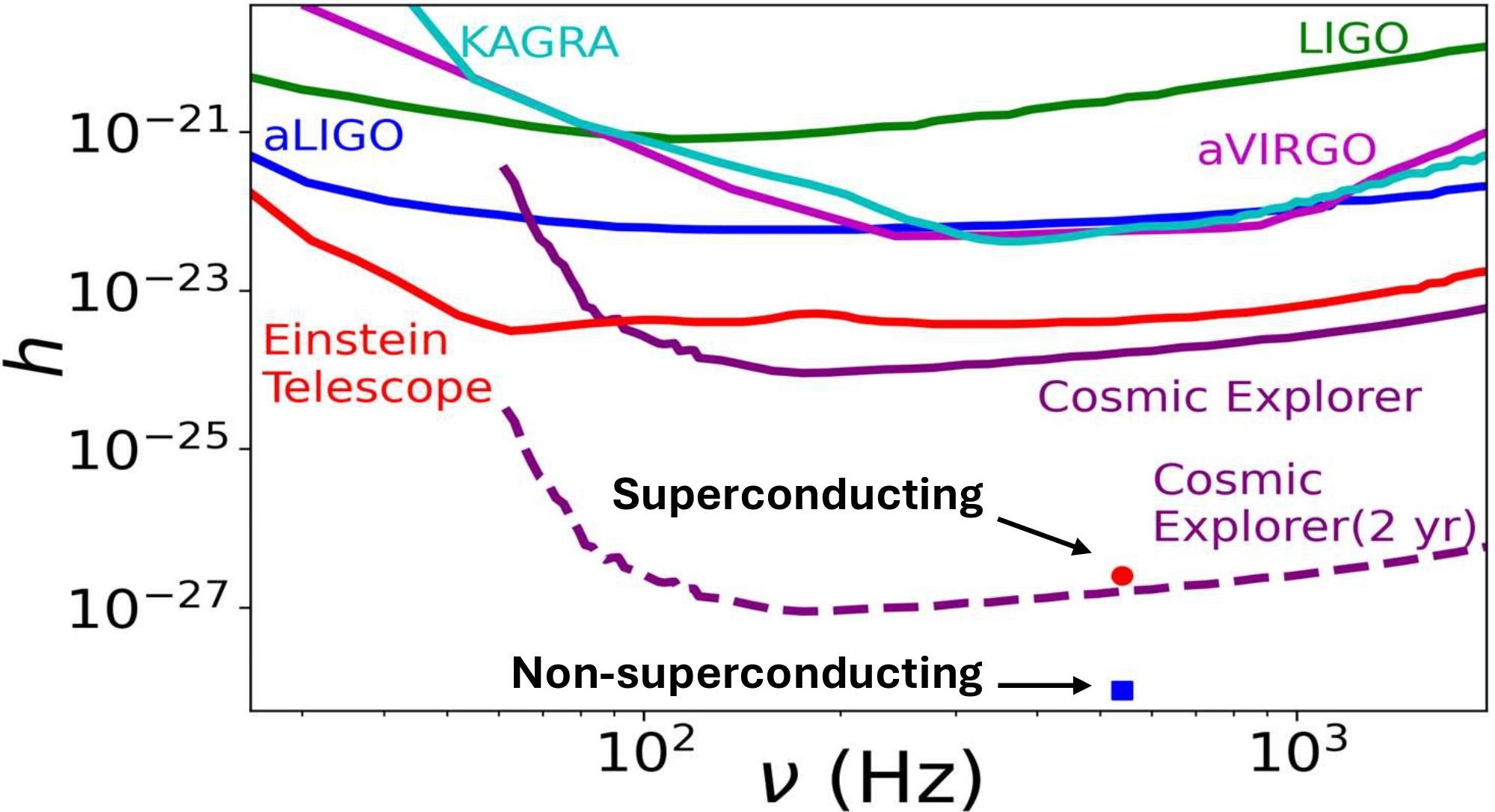}
\caption{ GW strain of MSP, along with the sensitivity of GW detectors.}
\label{fig:mspnew}
\end{center}
\end{figure}

A promising way to obtain indirect evidence for superconductivity is by searching for continuous gravitational waves (CGWs) emitted by isolated triaxial millisecond pulsars (MSPs), where the magnetic and rotational axes are misaligned~\cite{das-mukhopadhyay}. Rapidly spinning MSPs may have much stronger internal fields than the observed surface dipolar field because they may be buried by accretion~\cite{Haskell18}. The toroidal field component below the NS surface could be one order of magnitude stronger than the surface dipolar field for stability, and the core field could further increase by another order of magnitude. Additionally, if the MSP's field is concealed, the toroidal component inside might be up to $2 \times 10^4$ times stronger than the observed dipolar field, see Fig.~\ref{fig:2}. Therefore, the internal field and underlying deformation can strongly influence the amplitude of the CGW signal. 
While, ellipticity parameter $\epsilon \propto B_S^2$ in normal matter \cite{easson1977,freibenrezz2012}, in a superconducting NSs with $H < H_{c1}\sim10^{15}$ G, it scales by an additional factor of $H_{c1} / H$ due to enhanced magnetic stress by array of fluxtubes, where fields are intense and localized~\cite{easson1977}. Consequently, deformation and CGW amplitude are higher in superconducting regions. We calculate the GW strain $h$ using the mass, radius, moment of inertia $I$, and $\epsilon$ obtained from XNS by modeling MSP PSR J1843-1113. Based on the detection of CGWs, $h$ may indicate whether the magnetar consists of normal or superconducting matter. For a superconducting NS, the strain $h$ for continuous search over time $t\sim 2$ year is given by~\cite{sold2021,das-mukhopadhyay}:
\begin{equation}
    h \sim 10^{-2} \left(\frac{11.4}{\sqrt{t}}\right) \frac{4G}{c^4} \frac{(2\pi\nu)^2 \epsilon I}{d}\frac{H_{c1}}{{\langle B_S\rangle}}.
    \label{eq:gwh}
\end{equation}
PSR J1843-1113 has at the surface $B_S=1.6\times10^9$ G, rotation frequency $\nu=540$ Hz, and resides at a distance of $d=1.26$ kpc 
\cite{fermilat}. We calculate its CGW amplitude by constructing a model with $M=1.4M_\odot$ and $B_{S \max}=3\times 10^{13}$ G using XNS code. We extract $I=5\times10^{44}$ g cm\textsuperscript{2}, $\epsilon=5\times 10^{-15}$ from the model. The average field $\langle B_S\rangle \sim B_{S \max}$ as the toroidal magnetic field is maximum near the mid-radius (e.g., outer core) of the
star, see Fig.~\ref{fig:2}. With such a field,  a superconducting NS  will generate GW amplitude, which is at least two orders of magnitude higher ($h=10^{-27}$) than its non-superconducting counterpart ($h=3\times10^{-29}$), as shown in Fig.~\ref{fig:mspnew}. {If the detection of CGW from PSR J1843-1113 is feasible with the upcoming Cosmic Explorer
over a two-year observation period, it can serve as evidence for a hidden $H$-field and superconductivity in MSP.}

\section{Conclusions}
\label{sec:Conclusions}

We have explored the location of the type-II and type-I regions in magnetar models having toroidal fields,
with various masses, temperatures, and maximal values of the magnetic field using XNS code.
In each case, we have provided 1D and 2D representations of the volume and topology occupied by type-II and type-I regions
and their states. We find that models with larger masses have narrower superconducting regions. In a strong $H$-field, superconductivity is absent unless its magnitude intersects with $H_{c2}$. An intersection with only
$H_{c2}$ forms a type-II fluxtube state near the crust-core interface. In contrast, 
 intersections with only $H_{c1}$ give rise to two type-II phases -- one located on the inner side: type-II Meissner state, and the other outside: type-II fluxtube state. This new feature is characteristic of toroidal fields. For weak fields below $H_{c1}$, the Meissner state is favored. Increasing temperature reduces the superconducting regions without changing phase arrangements controlled by the relative magnitude of fields. In type-I superconductors, decreasing the field produces a Meissner or layered-domain phase extending inward up to an unpaired region where $\Delta = 0$. The most prominent features that emerge in 2D are the change in the topology of the type-II region from an open at the poles ellipsoid in the high-field star to a full ellipsoid in the lower-field star and the prolate type-I regions, most notably seen in intermediate-mass stars.

 Our investigation has uncovered that stable models of magnetars contain a non-superconducting,  torus-shaped region.   Moreover, the inner core of all models is hardly superconducting unless their density and/or local $H$-field is quite low, leading
 to $M\lesssim M_\odot$ only. This argues against the common belief that {most neutron stars are superconducting down to their center.}  
Finally, we have explored the potential for the indirect detection of superconductivity signatures through CGWs emitted by MSPs. If Cosmic Explorer detects GWs from such MSPs, it could provide strong evidence for the presence of hidden magnetic fields within magnetars and NSs, offering a novel insight into the role of superconductivity in these enigmatic objects.

\section*{Acknowledgments} We thank Zenia Zuraiq (IISc) for discussions. Thanks are also due to Soumallya Mitra for technical support and the referee for very useful suggestions. M.\,D. acknowledges the Prime Minister’s Research Fellows (PMRF) scheme, with Ref. No TF/PMRF-22-5442.03. B.\,M. acknowledges a project funded by SERB, India, with Ref. No. CRG/2022/003460, for partial support towards this research. A.\,S. acknowledges support through Deutsche Forschungsgemeinschaft Grant
No. SE 1836/6-1 and the Polish NCN Grant No. 2023/51/B/ST9/02798. 

\bibliographystyle{JHEP} 
\bibliography{references,PPNP_mag} 

\begin{thebibliography}{45}%
\makeatletter
\providecommand \@ifxundefined [1]{%
 \@ifx{#1\undefined}
}%
\providecommand \@ifnum [1]{%
 \ifnum #1\expandafter \@firstoftwo
 \else \expandafter \@secondoftwo
 \fi
}%
\providecommand \@ifx [1]{%
 \ifx #1\expandafter \@firstoftwo
 \else \expandafter \@secondoftwo
 \fi
}%
\providecommand \natexlab [1]{#1}%
\providecommand \enquote  [1]{``#1''}%
\providecommand \bibnamefont  [1]{#1}%
\providecommand \bibfnamefont [1]{#1}%
\providecommand \citenamefont [1]{#1}%
\providecommand \href@noop [0]{\@secondoftwo}%
\providecommand \href [0]{\begingroup \@sanitize@url \@href}%
\providecommand \@href[1]{\@@startlink{#1}\@@href}%
\providecommand \@@href[1]{\endgroup#1\@@endlink}%
\providecommand \@sanitize@url [0]{\catcode `\\12\catcode `\$12\catcode
  `\&12\catcode `\#12\catcode `\^12\catcode `\_12\catcode `\%12\relax}%
\providecommand \@@startlink[1]{}%
\providecommand \@@endlink[0]{}%
\providecommand \url  [0]{\begingroup\@sanitize@url \@url }%
\providecommand \@url [1]{\endgroup\@href {#1}{\urlprefix }}%
\providecommand \urlprefix  [0]{URL }%
\providecommand \Eprint [0]{\href }%
\providecommand \doibase [0]{https://doi.org/}%
\providecommand \selectlanguage [0]{\@gobble}%
\providecommand \bibinfo  [0]{\@secondoftwo}%
\providecommand \bibfield  [0]{\@secondoftwo}%
\providecommand \translation [1]{[#1]}%
\providecommand \BibitemOpen [0]{}%
\providecommand \bibitemStop [0]{}%
\providecommand \bibitemNoStop [0]{.\EOS\space}%
\providecommand \EOS [0]{\spacefactor3000\relax}%
\providecommand \BibitemShut  [1]{\csname bibitem#1\endcsname}%
\let\auto@bib@innerbib\@empty
\bibitem [{\citenamefont {{Page}}\ \emph {et~al.}(2006)\citenamefont {{Page}},
  \citenamefont {{Geppert}},\ and\ \citenamefont {{Weber}}}]{Page06}%
  \BibitemOpen
  \bibfield  {author} {\bibinfo {author} {\bibfnamefont {D.}~\bibnamefont
  {{Page}}}, \bibinfo {author} {\bibfnamefont {U.}~\bibnamefont {{Geppert}}},\
  and\ \bibinfo {author} {\bibfnamefont {F.}~\bibnamefont {{Weber}}},\ }\href
  {https://doi.org/10.1016/j.nuclphysa.2005.09.019} {\bibfield  {journal}
  {\bibinfo  {journal} {\nphysa}\ }\textbf {\bibinfo {volume} {777}},\ \bibinfo
  {pages} {497} (\bibinfo {year} {2006})},\ \Eprint
  {https://arxiv.org/abs/astro-ph/0508056} {arXiv:astro-ph/0508056 [astro-ph]}
  \BibitemShut {NoStop}%
\bibitem [{\citenamefont {{Page}}\ \emph {et~al.}(2011)\citenamefont {{Page}},
  \citenamefont {{Prakash}}, \citenamefont {{Lattimer}},\ and\ \citenamefont
  {{Steiner}}}]{Page2011PhRvL}%
  \BibitemOpen
  \bibfield  {author} {\bibinfo {author} {\bibfnamefont {D.}~\bibnamefont
  {{Page}}}, \bibinfo {author} {\bibfnamefont {M.}~\bibnamefont {{Prakash}}},
  \bibinfo {author} {\bibfnamefont {J.~M.}\ \bibnamefont {{Lattimer}}},\ and\
  \bibinfo {author} {\bibfnamefont {A.~W.}\ \bibnamefont {{Steiner}}},\ }\href
  {https://doi.org/10.1103/PhysRevLett.106.081101} {\bibfield  {journal}
  {\bibinfo  {journal} {\prl}\ }\textbf {\bibinfo {volume} {106}},\ \bibinfo
  {eid} {081101} (\bibinfo {year} {2011})},\ \Eprint
  {https://arxiv.org/abs/1011.6142} {arXiv:1011.6142 [astro-ph.HE]}
  \BibitemShut {NoStop}%
\bibitem [{\citenamefont {{Blaschke}}\ \emph {et~al.}(2013)\citenamefont
  {{Blaschke}}, \citenamefont {{Grigorian}},\ and\ \citenamefont
  {{Voskresensky}}}]{Blaschke2013PhRvC}%
  \BibitemOpen
  \bibfield  {author} {\bibinfo {author} {\bibfnamefont {D.}~\bibnamefont
  {{Blaschke}}}, \bibinfo {author} {\bibfnamefont {H.}~\bibnamefont
  {{Grigorian}}},\ and\ \bibinfo {author} {\bibfnamefont {D.~N.}\ \bibnamefont
  {{Voskresensky}}},\ }\href {https://doi.org/10.1103/PhysRevC.88.065805}
  {\bibfield  {journal} {\bibinfo  {journal} {\prc}\ }\textbf {\bibinfo
  {volume} {88}},\ \bibinfo {eid} {065805} (\bibinfo {year} {2013})},\ \Eprint
  {https://arxiv.org/abs/1308.4093} {arXiv:1308.4093 [nucl-th]} \BibitemShut
  {NoStop}%
\bibitem [{\citenamefont {Abrikosov}(1988)}]{Abrikosov:Fundamentals}%
  \BibitemOpen
  \bibfield  {author} {\bibinfo {author} {\bibfnamefont {A.}~\bibnamefont
  {Abrikosov}},\ }\href@noop {} {\emph {\bibinfo {title} {Fundamentals of the
  Theory of Metals}}}\ (\bibinfo  {publisher} {North-Holland},\ \bibinfo
  {address} {Amsterdam},\ \bibinfo {year} {1988})\BibitemShut {NoStop}%
\bibitem [{\citenamefont {{Sedrakian}}\ and\ \citenamefont
  {{Clark}}(2019)}]{Sedrakian2019EPJA}%
  \BibitemOpen
  \bibfield  {author} {\bibinfo {author} {\bibfnamefont {A.}~\bibnamefont
  {{Sedrakian}}}\ and\ \bibinfo {author} {\bibfnamefont {J.~W.}\ \bibnamefont
  {{Clark}}},\ }\href {https://doi.org/10.1140/epja/i2019-12863-6} {\bibfield
  {journal} {\bibinfo  {journal} {European Physical Journal A}\ }\textbf
  {\bibinfo {volume} {55}},\ \bibinfo {eid} {167} (\bibinfo {year} {2019})},\
  \Eprint {https://arxiv.org/abs/1802.00017} {arXiv:1802.00017 [nucl-th]}
  \BibitemShut {NoStop}%
\bibitem [{\citenamefont {{Chamel}}(2017)}]{Chamel2017}%
  \BibitemOpen
  \bibfield  {author} {\bibinfo {author} {\bibfnamefont {N.}~\bibnamefont
  {{Chamel}}},\ }\href {https://doi.org/10.1007/s12036-017-9470-9} {\bibfield
  {journal} {\bibinfo  {journal} {\aap}\ }\textbf {\bibinfo {volume} {38}},\
  \bibinfo {eid} {43} (\bibinfo {year} {2017})},\ \Eprint
  {https://arxiv.org/abs/1709.07288} {arXiv:1709.07288 [astro-ph.HE]}
  \BibitemShut {NoStop}%
\bibitem [{\citenamefont {{Haskell}}\ and\ \citenamefont
  {{Sedrakian}}(2018)}]{Haskell18}%
  \BibitemOpen
  \bibfield  {author} {\bibinfo {author} {\bibfnamefont {B.}~\bibnamefont
  {{Haskell}}}\ and\ \bibinfo {author} {\bibfnamefont {A.}~\bibnamefont
  {{Sedrakian}}},\ }in\ \href {https://doi.org/10.1007/978-3-319-97616-7_8}
  {\emph {\bibinfo {booktitle} {ASSL}}},\ Vol.\ \bibinfo {volume} {457}\
  (\bibinfo {year} {2018})\ p.\ \bibinfo {pages} {401},\ \Eprint
  {https://arxiv.org/abs/1709.10340} {arXiv:1709.10340 [astro-ph.HE]}
  \BibitemShut {NoStop}%
\bibitem [{\citenamefont {{Zhou}}\ \emph {et~al.}(2022)\citenamefont {{Zhou}},
  \citenamefont {{G{\"u}gercino{\u{g}}lu}}, \citenamefont {{Yuan}},
  \citenamefont {{Ge}},\ and\ \citenamefont {{Yu}}}]{GlitchesZhou2022}%
  \BibitemOpen
  \bibfield  {author} {\bibinfo {author} {\bibfnamefont {S.}~\bibnamefont
  {{Zhou}}}, \bibinfo {author} {\bibfnamefont {E.}~\bibnamefont
  {{G{\"u}gercino{\u{g}}lu}}}, \bibinfo {author} {\bibfnamefont
  {J.}~\bibnamefont {{Yuan}}}, \bibinfo {author} {\bibfnamefont
  {M.}~\bibnamefont {{Ge}}},\ and\ \bibinfo {author} {\bibfnamefont
  {C.}~\bibnamefont {{Yu}}},\ }\href {https://doi.org/10.3390/universe8120641}
  {\bibfield  {journal} {\bibinfo  {journal} {Universe}\ }\textbf {\bibinfo
  {volume} {8}},\ \bibinfo {eid} {641} (\bibinfo {year} {2022})},\ \Eprint
  {https://arxiv.org/abs/2211.13885} {arXiv:2211.13885 [astro-ph.HE]}
  \BibitemShut {NoStop}%
\bibitem [{\citenamefont {{Antonopoulou}}\ \emph {et~al.}(2022)\citenamefont
  {{Antonopoulou}}, \citenamefont {{Haskell}},\ and\ \citenamefont
  {{Espinoza}}}]{Antonopoulou2022RPPh}%
  \BibitemOpen
  \bibfield  {author} {\bibinfo {author} {\bibfnamefont {D.}~\bibnamefont
  {{Antonopoulou}}}, \bibinfo {author} {\bibfnamefont {B.}~\bibnamefont
  {{Haskell}}},\ and\ \bibinfo {author} {\bibfnamefont {C.~M.}\ \bibnamefont
  {{Espinoza}}},\ }\href {https://doi.org/10.1088/1361-6633/ac9ced} {\bibfield
  {journal} {\bibinfo  {journal} {Reports on Progress in Physics}\ }\textbf
  {\bibinfo {volume} {85}},\ \bibinfo {eid} {126901} (\bibinfo {year}
  {2022})}\BibitemShut {NoStop}%
\bibitem [{\citenamefont {{Pons}}\ \emph {et~al.}(2009)\citenamefont {{Pons}},
  \citenamefont {{Miralles}},\ and\ \citenamefont {{Geppert}}}]{Pons2009AA}%
  \BibitemOpen
  \bibfield  {author} {\bibinfo {author} {\bibfnamefont {J.~A.}\ \bibnamefont
  {{Pons}}}, \bibinfo {author} {\bibfnamefont {J.~A.}\ \bibnamefont
  {{Miralles}}},\ and\ \bibinfo {author} {\bibfnamefont {U.}~\bibnamefont
  {{Geppert}}},\ }\href {https://doi.org/10.1051/0004-6361:200811229}
  {\bibfield  {journal} {\bibinfo  {journal} {\aap}\ }\textbf {\bibinfo
  {volume} {496}},\ \bibinfo {pages} {207} (\bibinfo {year} {2009})},\ \Eprint
  {https://arxiv.org/abs/0812.3018} {arXiv:0812.3018 [astro-ph]} \BibitemShut
  {NoStop}%
\bibitem [{\citenamefont {{Ascenzi}}\ \emph {et~al.}(2024)\citenamefont
  {{Ascenzi}}, \citenamefont {{Vigano}}, \citenamefont {{Dehman}},
  \citenamefont {{Pons}}, \citenamefont {{Rea}},\ and\ \citenamefont
  {{Perna}}}]{Ascenzi2024MNRAS}%
  \BibitemOpen
  \bibfield  {author} {\bibinfo {author} {\bibfnamefont {S.}~\bibnamefont
  {{Ascenzi}}}, \bibinfo {author} {\bibfnamefont {D.}~\bibnamefont {{Vigano}}},
  \bibinfo {author} {\bibfnamefont {C.}~\bibnamefont {{Dehman}}}, \bibinfo
  {author} {\bibfnamefont {J.~A.}\ \bibnamefont {{Pons}}}, \bibinfo {author}
  {\bibfnamefont {N.}~\bibnamefont {{Rea}}},\ and\ \bibinfo {author}
  {\bibfnamefont {R.}~\bibnamefont {{Perna}}},\ }\bibfield  {journal} {\bibinfo
   {journal} {\mnras}\ }\href {https://doi.org/10.1093/mnras/stae1749}
  {10.1093/mnras/stae1749} (\bibinfo {year} {2024}),\ \Eprint
  {https://arxiv.org/abs/2401.15711} {arXiv:2401.15711 [astro-ph.HE]}
  \BibitemShut {NoStop}%
\bibitem [{\citenamefont {{Alford}}\ and\ \citenamefont
  {{Schwenzer}}(2014)}]{Alford2014ApJ}%
  \BibitemOpen
  \bibfield  {author} {\bibinfo {author} {\bibfnamefont {M.~G.}\ \bibnamefont
  {{Alford}}}\ and\ \bibinfo {author} {\bibfnamefont {K.}~\bibnamefont
  {{Schwenzer}}},\ }\href {https://doi.org/10.1088/0004-637X/781/1/26}
  {\bibfield  {journal} {\bibinfo  {journal} {\apj}\ }\textbf {\bibinfo
  {volume} {781}},\ \bibinfo {eid} {26} (\bibinfo {year} {2014})},\ \Eprint
  {https://arxiv.org/abs/1210.6091} {arXiv:1210.6091 [gr-qc]} \BibitemShut
  {NoStop}%
\bibitem [{\citenamefont {{Piccinni}}(2022)}]{Piccinni2022}%
  \BibitemOpen
  \bibfield  {author} {\bibinfo {author} {\bibfnamefont {O.~J.}\ \bibnamefont
  {{Piccinni}}},\ }\href {https://doi.org/10.3390/galaxies10030072} {\bibfield
  {journal} {\bibinfo  {journal} {Galaxies}\ }\textbf {\bibinfo {volume}
  {10}},\ \bibinfo {eid} {72} (\bibinfo {year} {2022})},\ \Eprint
  {https://arxiv.org/abs/2202.01088} {arXiv:2202.01088 [gr-qc]} \BibitemShut
  {NoStop}%
\bibitem [{\citenamefont {{Andersson}}(2021)}]{Andersson2021Univ}%
  \BibitemOpen
  \bibfield  {author} {\bibinfo {author} {\bibfnamefont {N.}~\bibnamefont
  {{Andersson}}},\ }\href {https://doi.org/10.3390/universe7010017} {\bibfield
  {journal} {\bibinfo  {journal} {Universe}\ }\textbf {\bibinfo {volume} {7}},\
  \bibinfo {eid} {17} (\bibinfo {year} {2021})},\ \Eprint
  {https://arxiv.org/abs/2103.10218} {arXiv:2103.10218 [astro-ph.HE]}
  \BibitemShut {NoStop}%
\bibitem [{\citenamefont {{Graber}}\ \emph {et~al.}(2017)\citenamefont
  {{Graber}}, \citenamefont {{Andersson}},\ and\ \citenamefont
  {{Hogg}}}]{Graber2017}%
  \BibitemOpen
  \bibfield  {author} {\bibinfo {author} {\bibfnamefont {V.}~\bibnamefont
  {{Graber}}}, \bibinfo {author} {\bibfnamefont {N.}~\bibnamefont
  {{Andersson}}},\ and\ \bibinfo {author} {\bibfnamefont {M.}~\bibnamefont
  {{Hogg}}},\ }\href {https://doi.org/10.1142/S0218271817300154} {\bibfield
  {journal} {\bibinfo  {journal} {International Journal of Modern Physics D}\
  }\textbf {\bibinfo {volume} {26}},\ \bibinfo {eid} {1730015-347} (\bibinfo
  {year} {2017})},\ \Eprint {https://arxiv.org/abs/1610.06882}
  {arXiv:1610.06882 [astro-ph.HE]} \BibitemShut {NoStop}%
\bibitem [{\citenamefont {{Stein}}\ \emph {et~al.}(2016)\citenamefont
  {{Stein}}, \citenamefont {{Sedrakian}}, \citenamefont {{Huang}},\ and\
  \citenamefont {{Clark}}}]{Stein2016PhRvC}%
  \BibitemOpen
  \bibfield  {author} {\bibinfo {author} {\bibfnamefont {M.}~\bibnamefont
  {{Stein}}}, \bibinfo {author} {\bibfnamefont {A.}~\bibnamefont
  {{Sedrakian}}}, \bibinfo {author} {\bibfnamefont {X.-G.}\ \bibnamefont
  {{Huang}}},\ and\ \bibinfo {author} {\bibfnamefont {J.~W.}\ \bibnamefont
  {{Clark}}},\ }\href {https://doi.org/10.1103/PhysRevC.93.015802} {\bibfield
  {journal} {\bibinfo  {journal} {\prc}\ }\textbf {\bibinfo {volume} {93}},\
  \bibinfo {eid} {015802} (\bibinfo {year} {2016})},\ \Eprint
  {https://arxiv.org/abs/1510.06000} {arXiv:1510.06000 [nucl-th]} \BibitemShut
  {NoStop}%
\bibitem [{\citenamefont {{Sinha}}\ and\ \citenamefont
  {{Sedrakian}}(2015{\natexlab{a}})}]{Sinha2015PhRvC}%
  \BibitemOpen
  \bibfield  {author} {\bibinfo {author} {\bibfnamefont {M.}~\bibnamefont
  {{Sinha}}}\ and\ \bibinfo {author} {\bibfnamefont {A.}~\bibnamefont
  {{Sedrakian}}},\ }\href {https://doi.org/10.1103/PhysRevC.91.035805}
  {\bibfield  {journal} {\bibinfo  {journal} {\prc}\ }\textbf {\bibinfo
  {volume} {91}},\ \bibinfo {eid} {035805} (\bibinfo {year}
  {2015}{\natexlab{a}})},\ \Eprint {https://arxiv.org/abs/1502.02979}
  {arXiv:1502.02979 [astro-ph.HE]} \BibitemShut {NoStop}%
\bibitem [{\citenamefont {{Haber}}\ and\ \citenamefont
  {{Schmitt}}(2017)}]{Haber2017PhRvD}%
  \BibitemOpen
  \bibfield  {author} {\bibinfo {author} {\bibfnamefont {A.}~\bibnamefont
  {{Haber}}}\ and\ \bibinfo {author} {\bibfnamefont {A.}~\bibnamefont
  {{Schmitt}}},\ }\href {https://doi.org/10.1103/PhysRevD.95.116016} {\bibfield
   {journal} {\bibinfo  {journal} {\prd}\ }\textbf {\bibinfo {volume} {95}},\
  \bibinfo {eid} {116016} (\bibinfo {year} {2017})},\ \Eprint
  {https://arxiv.org/abs/1704.01575} {arXiv:1704.01575 [hep-th]} \BibitemShut
  {NoStop}%
\bibitem [{\citenamefont {{Sinha}}\ and\ \citenamefont
  {{Sedrakian}}(2015{\natexlab{b}})}]{Sinha2015PPN}%
  \BibitemOpen
  \bibfield  {author} {\bibinfo {author} {\bibfnamefont {M.}~\bibnamefont
  {{Sinha}}}\ and\ \bibinfo {author} {\bibfnamefont {A.}~\bibnamefont
  {{Sedrakian}}},\ }\href {https://doi.org/10.1134/S1063779615050275}
  {\bibfield  {journal} {\bibinfo  {journal} {Physics of Particles and Nuclei}\
  }\textbf {\bibinfo {volume} {46}},\ \bibinfo {pages} {826} (\bibinfo {year}
  {2015}{\natexlab{b}})},\ \Eprint {https://arxiv.org/abs/1403.2829}
  {arXiv:1403.2829 [astro-ph.SR]} \BibitemShut {NoStop}%
\bibitem [{\citenamefont {{Chatterjee}}\ \emph {et~al.}(2019)\citenamefont
  {{Chatterjee}}, \citenamefont {{Novak}},\ and\ \citenamefont
  {{Oertel}}}]{Chatterjee2019PhRvC}%
  \BibitemOpen
  \bibfield  {author} {\bibinfo {author} {\bibfnamefont {D.}~\bibnamefont
  {{Chatterjee}}}, \bibinfo {author} {\bibfnamefont {J.}~\bibnamefont
  {{Novak}}},\ and\ \bibinfo {author} {\bibfnamefont {M.}~\bibnamefont
  {{Oertel}}},\ }\href {https://doi.org/10.1103/PhysRevC.99.055811} {\bibfield
  {journal} {\bibinfo  {journal} {\prc}\ }\textbf {\bibinfo {volume} {99}},\
  \bibinfo {eid} {055811} (\bibinfo {year} {2019})},\ \Eprint
  {https://arxiv.org/abs/1808.01778} {arXiv:1808.01778 [nucl-th]} \BibitemShut
  {NoStop}%
\bibitem [{\citenamefont {{Raduta}}\ \emph {et~al.}(2019)\citenamefont
  {{Raduta}}, \citenamefont {{Li}}, \citenamefont {{Sedrakian}},\ and\
  \citenamefont {{Weber}}}]{Raduta2019MNRAS}%
  \BibitemOpen
  \bibfield  {author} {\bibinfo {author} {\bibfnamefont {A.~R.}\ \bibnamefont
  {{Raduta}}}, \bibinfo {author} {\bibfnamefont {J.~J.}\ \bibnamefont {{Li}}},
  \bibinfo {author} {\bibfnamefont {A.}~\bibnamefont {{Sedrakian}}},\ and\
  \bibinfo {author} {\bibfnamefont {F.}~\bibnamefont {{Weber}}},\ }\href
  {https://doi.org/10.1093/mnras/stz1459} {\bibfield  {journal} {\bibinfo
  {journal} {\mnras}\ }\textbf {\bibinfo {volume} {487}},\ \bibinfo {pages}
  {2639} (\bibinfo {year} {2019})},\ \Eprint {https://arxiv.org/abs/1903.01295}
  {arXiv:1903.01295 [nucl-th]} \BibitemShut {NoStop}%
\bibitem [{\citenamefont {{Alm}}\ \emph {et~al.}(1996)\citenamefont {{Alm}},
  \citenamefont {{R{\"o}pke}}, \citenamefont {{Sedrakian}},\ and\ \citenamefont
  {{Weber}}}]{Alm1996NuPhA}%
  \BibitemOpen
  \bibfield  {author} {\bibinfo {author} {\bibfnamefont {T.}~\bibnamefont
  {{Alm}}}, \bibinfo {author} {\bibfnamefont {G.}~\bibnamefont {{R{\"o}pke}}},
  \bibinfo {author} {\bibfnamefont {A.}~\bibnamefont {{Sedrakian}}},\ and\
  \bibinfo {author} {\bibfnamefont {F.}~\bibnamefont {{Weber}}},\ }\href
  {https://doi.org/10.1016/0375-9474(96)00153-4} {\bibfield  {journal}
  {\bibinfo  {journal} {\nphysa}\ }\textbf {\bibinfo {volume} {604}},\ \bibinfo
  {pages} {491} (\bibinfo {year} {1996})}\BibitemShut {NoStop}%
\bibitem [{\citenamefont {{Soldateschi}}\ \emph {et~al.}(2021)\citenamefont
  {{Soldateschi}}, \citenamefont {{Bucciantini}},\ and\ \citenamefont {{Del
  Zanna}}}]{sold2021main}%
  \BibitemOpen
  \bibfield  {author} {\bibinfo {author} {\bibfnamefont {J.}~\bibnamefont
  {{Soldateschi}}}, \bibinfo {author} {\bibfnamefont {N.}~\bibnamefont
  {{Bucciantini}}},\ and\ \bibinfo {author} {\bibfnamefont {L.}~\bibnamefont
  {{Del Zanna}}},\ }\href {https://doi.org/10.1051/0004-6361/202141448}
  {\bibfield  {journal} {\bibinfo  {journal} {\aap}\ }\textbf {\bibinfo
  {volume} {654}},\ \bibinfo {eid} {A162} (\bibinfo {year} {2021})},\ \Eprint
  {https://arxiv.org/abs/2106.00603} {arXiv:2106.00603 [astro-ph.HE]}
  \BibitemShut {NoStop}%
\bibitem [{\citenamefont {{Akg{\"u}n}}\ and\ \citenamefont
  {{Wasserman}}(2008)}]{AW2008MNRAS}%
  \BibitemOpen
  \bibfield  {author} {\bibinfo {author} {\bibfnamefont {T.}~\bibnamefont
  {{Akg{\"u}n}}}\ and\ \bibinfo {author} {\bibfnamefont {I.}~\bibnamefont
  {{Wasserman}}},\ }\href {https://doi.org/10.1111/j.1365-2966.2007.12660.x}
  {\bibfield  {journal} {\bibinfo  {journal} {\mnras}\ }\textbf {\bibinfo
  {volume} {383}},\ \bibinfo {pages} {1551} (\bibinfo {year} {2008})},\ \Eprint
  {https://arxiv.org/abs/0705.2195} {arXiv:0705.2195 [astro-ph]} \BibitemShut
  {NoStop}%
\bibitem [{\citenamefont {{Lander}}\ \emph {et~al.}(2012)\citenamefont
  {{Lander}}, \citenamefont {{Andersson}},\ and\ \citenamefont
  {{Glampedakis}}}]{Lander12}%
  \BibitemOpen
  \bibfield  {author} {\bibinfo {author} {\bibfnamefont {S.~K.}\ \bibnamefont
  {{Lander}}}, \bibinfo {author} {\bibfnamefont {N.}~\bibnamefont
  {{Andersson}}},\ and\ \bibinfo {author} {\bibfnamefont {K.}~\bibnamefont
  {{Glampedakis}}},\ }\href {https://doi.org/10.1111/j.1365-2966.2011.19720.x}
  {\bibfield  {journal} {\bibinfo  {journal} {\mnras}\ }\textbf {\bibinfo
  {volume} {419}},\ \bibinfo {pages} {732} (\bibinfo {year} {2012})},\ \Eprint
  {https://arxiv.org/abs/1106.6322} {arXiv:1106.6322 [astro-ph.SR]}
  \BibitemShut {NoStop}%
\bibitem [{\citenamefont {{M{\"u}hlschlegel}}(1959)}]{Muhlschlegel1959ZPhy}%
  \BibitemOpen
  \bibfield  {author} {\bibinfo {author} {\bibfnamefont {B.}~\bibnamefont
  {{M{\"u}hlschlegel}}},\ }\href {https://doi.org/10.1007/BF01332932}
  {\bibfield  {journal} {\bibinfo  {journal} {Zeitschrift fur Physik}\ }\textbf
  {\bibinfo {volume} {155}},\ \bibinfo {pages} {313} (\bibinfo {year}
  {1959})}\BibitemShut {NoStop}%
\bibitem [{\citenamefont {{Grill}}\ \emph {et~al.}(2014)\citenamefont
  {{Grill}}, \citenamefont {{Pais}}, \citenamefont {{Provid{\^e}ncia}},
  \citenamefont {{Vida{\~n}a}},\ and\ \citenamefont {{Avancini}}}]{gppva}%
  \BibitemOpen
  \bibfield  {author} {\bibinfo {author} {\bibfnamefont {F.}~\bibnamefont
  {{Grill}}}, \bibinfo {author} {\bibfnamefont {H.}~\bibnamefont {{Pais}}},
  \bibinfo {author} {\bibfnamefont {C.}~\bibnamefont {{Provid{\^e}ncia}}},
  \bibinfo {author} {\bibfnamefont {I.}~\bibnamefont {{Vida{\~n}a}}},\ and\
  \bibinfo {author} {\bibfnamefont {S.~S.}\ \bibnamefont {{Avancini}}},\ }\href
  {https://doi.org/10.1103/PhysRevC.90.045803} {\bibfield  {journal} {\bibinfo
  {journal} {\prc}\ }\textbf {\bibinfo {volume} {90}},\ \bibinfo {eid} {045803}
  (\bibinfo {year} {2014})},\ \Eprint {https://arxiv.org/abs/1404.2753}
  {arXiv:1404.2753 [nucl-th]} \BibitemShut {NoStop}%
\bibitem [{\citenamefont {{Lalazissis}}\ \emph {et~al.}(2005)\citenamefont
  {{Lalazissis}}, \citenamefont {{Nik{\v{s}}i{\'c}}}, \citenamefont
  {{Vretenar}},\ and\ \citenamefont {{Ring}}}]{Lala2005}%
  \BibitemOpen
  \bibfield  {author} {\bibinfo {author} {\bibfnamefont {G.~A.}\ \bibnamefont
  {{Lalazissis}}}, \bibinfo {author} {\bibfnamefont {T.}~\bibnamefont
  {{Nik{\v{s}}i{\'c}}}}, \bibinfo {author} {\bibfnamefont {D.}~\bibnamefont
  {{Vretenar}}},\ and\ \bibinfo {author} {\bibfnamefont {P.}~\bibnamefont
  {{Ring}}},\ }\href {https://doi.org/10.1103/PhysRevC.71.024312} {\bibfield
  {journal} {\bibinfo  {journal} {\prc}\ }\textbf {\bibinfo {volume} {71}},\
  \bibinfo {eid} {024312} (\bibinfo {year} {2005})}\BibitemShut {NoStop}%
\bibitem [{\citenamefont {{CompOSE Core Team}}\ \emph
  {et~al.}(2022)\citenamefont {{CompOSE Core Team}}, \citenamefont {{Typel}},
  \citenamefont {{Oertel}}, \citenamefont {{Kl{\"a}hn}}, \citenamefont
  {{Chatterjee}}, \citenamefont {{Dexheimer}}, \citenamefont {{Ishizuka}},
  \citenamefont {{Mancini}}, \citenamefont {{Novak}}, \citenamefont {{Pais}},
  \citenamefont {{Provid{\^e}ncia}}, \citenamefont {{R. Raduta}}, \citenamefont
  {{Servillat}},\ and\ \citenamefont {{Tolos}}}]{CompOSE2022EPJA}%
  \BibitemOpen
  \bibfield  {author} {\bibinfo {author} {\bibnamefont {{CompOSE Core Team}}},
  \bibinfo {author} {\bibfnamefont {S.}~\bibnamefont {{Typel}}}, \bibinfo
  {author} {\bibfnamefont {M.}~\bibnamefont {{Oertel}}}, \bibinfo {author}
  {\bibfnamefont {T.}~\bibnamefont {{Kl{\"a}hn}}}, \bibinfo {author}
  {\bibfnamefont {D.}~\bibnamefont {{Chatterjee}}}, \bibinfo {author}
  {\bibfnamefont {V.}~\bibnamefont {{Dexheimer}}}, \bibinfo {author}
  {\bibfnamefont {C.}~\bibnamefont {{Ishizuka}}}, \bibinfo {author}
  {\bibfnamefont {M.}~\bibnamefont {{Mancini}}}, \bibinfo {author}
  {\bibfnamefont {J.}~\bibnamefont {{Novak}}}, \bibinfo {author} {\bibfnamefont
  {H.}~\bibnamefont {{Pais}}}, \bibinfo {author} {\bibfnamefont
  {C.}~\bibnamefont {{Provid{\^e}ncia}}}, \bibinfo {author} {\bibfnamefont
  {A.}~\bibnamefont {{R. Raduta}}}, \bibinfo {author} {\bibfnamefont
  {M.}~\bibnamefont {{Servillat}}},\ and\ \bibinfo {author} {\bibfnamefont
  {L.}~\bibnamefont {{Tolos}}},\ }\href
  {https://doi.org/10.1140/epja/s10050-022-00847-y} {\bibfield  {journal}
  {\bibinfo  {journal} {European Physical Journal A}\ }\textbf {\bibinfo
  {volume} {58}},\ \bibinfo {eid} {221} (\bibinfo {year} {2022})},\ \Eprint
  {https://arxiv.org/abs/2203.03209} {arXiv:2203.03209 [astro-ph.HE]}
  \BibitemShut {NoStop}%
\bibitem [{\citenamefont {Dexheimer}\ \emph {et~al.}(2022)\citenamefont
  {Dexheimer}, \citenamefont {Mancini}, \citenamefont {Oertel}, \citenamefont
  {Providência}, \citenamefont {Tolos},\ and\ \citenamefont
  {Typel}}]{Dexheimer2022}%
  \BibitemOpen
  \bibfield  {author} {\bibinfo {author} {\bibfnamefont {V.}~\bibnamefont
  {Dexheimer}}, \bibinfo {author} {\bibfnamefont {M.}~\bibnamefont {Mancini}},
  \bibinfo {author} {\bibfnamefont {M.}~\bibnamefont {Oertel}}, \bibinfo
  {author} {\bibfnamefont {C.}~\bibnamefont {Providência}}, \bibinfo {author}
  {\bibfnamefont {L.}~\bibnamefont {Tolos}},\ and\ \bibinfo {author}
  {\bibfnamefont {S.}~\bibnamefont {Typel}},\ }\href
  {https://doi.org/10.3390/particles5030028} {\bibfield  {journal} {\bibinfo
  {journal} {Particles}\ }\textbf {\bibinfo {volume} {5}},\ \bibinfo {pages}
  {346} (\bibinfo {year} {2022})}\BibitemShut {NoStop}%
\bibitem [{\citenamefont {{Sedrakian}}\ and\ \citenamefont
  {{Sedrakian}}(1995)}]{Sedrakian1995ApJ}%
  \BibitemOpen
  \bibfield  {author} {\bibinfo {author} {\bibfnamefont {A.~D.}\ \bibnamefont
  {{Sedrakian}}}\ and\ \bibinfo {author} {\bibfnamefont {D.~M.}\ \bibnamefont
  {{Sedrakian}}},\ }\href {https://doi.org/10.1086/175876} {\bibfield
  {journal} {\bibinfo  {journal} {\apj}\ }\textbf {\bibinfo {volume} {447}},\
  \bibinfo {pages} {305} (\bibinfo {year} {1995})}\BibitemShut {NoStop}%
\bibitem [{\citenamefont {{Baym}}\ \emph {et~al.}(1969)\citenamefont {{Baym}},
  \citenamefont {{Pethick}},\ and\ \citenamefont {{Pines}}}]{Baym1969}%
  \BibitemOpen
  \bibfield  {author} {\bibinfo {author} {\bibfnamefont {G.}~\bibnamefont
  {{Baym}}}, \bibinfo {author} {\bibfnamefont {C.}~\bibnamefont {{Pethick}}},\
  and\ \bibinfo {author} {\bibfnamefont {D.}~\bibnamefont {{Pines}}},\ }\href
  {https://doi.org/10.1038/224673a0} {\bibfield  {journal} {\bibinfo  {journal}
  {\nat}\ }\textbf {\bibinfo {volume} {224}},\ \bibinfo {pages} {673} (\bibinfo
  {year} {1969})}\BibitemShut {NoStop}%
\bibitem [{\citenamefont {{Glampedakis}}\ \emph {et~al.}(2011)\citenamefont
  {{Glampedakis}}, \citenamefont {{Andersson}},\ and\ \citenamefont
  {{Samuelsson}}}]{Glampedakis2011}%
  \BibitemOpen
  \bibfield  {author} {\bibinfo {author} {\bibfnamefont {K.}~\bibnamefont
  {{Glampedakis}}}, \bibinfo {author} {\bibfnamefont {N.}~\bibnamefont
  {{Andersson}}},\ and\ \bibinfo {author} {\bibfnamefont {L.}~\bibnamefont
  {{Samuelsson}}},\ }\href {https://doi.org/10.1111/j.1365-2966.2010.17484.x}
  {\bibfield  {journal} {\bibinfo  {journal} {\mnras}\ }\textbf {\bibinfo
  {volume} {410}},\ \bibinfo {pages} {805} (\bibinfo {year} {2011})},\ \Eprint
  {https://arxiv.org/abs/1001.4046} {arXiv:1001.4046 [astro-ph.SR]}
  \BibitemShut {NoStop}%
\bibitem [{\citenamefont {{Gusakov}}\ and\ \citenamefont
  {{Dommes}}(2016)}]{Gusakov2016a}%
  \BibitemOpen
  \bibfield  {author} {\bibinfo {author} {\bibfnamefont {M.~E.}\ \bibnamefont
  {{Gusakov}}}\ and\ \bibinfo {author} {\bibfnamefont {V.~A.}\ \bibnamefont
  {{Dommes}}},\ }\href {https://doi.org/10.1103/PhysRevD.94.083006} {\bibfield
  {journal} {\bibinfo  {journal} {\prd}\ }\textbf {\bibinfo {volume} {94}},\
  \bibinfo {eid} {083006} (\bibinfo {year} {2016})},\ \Eprint
  {https://arxiv.org/abs/1607.01629} {arXiv:1607.01629 [gr-qc]} \BibitemShut
  {NoStop}%
\bibitem [{\citenamefont {{Rau}}\ and\ \citenamefont
  {{Wasserman}}(2020)}]{Rau2020}%
  \BibitemOpen
  \bibfield  {author} {\bibinfo {author} {\bibfnamefont {P.~B.}\ \bibnamefont
  {{Rau}}}\ and\ \bibinfo {author} {\bibfnamefont {I.}~\bibnamefont
  {{Wasserman}}},\ }\href {https://doi.org/10.1103/PhysRevD.102.063011}
  {\bibfield  {journal} {\bibinfo  {journal} {\prd}\ }\textbf {\bibinfo
  {volume} {102}},\ \bibinfo {eid} {063011} (\bibinfo {year} {2020})},\ \Eprint
  {https://arxiv.org/abs/2004.07468} {arXiv:2004.07468 [astro-ph.HE]}
  \BibitemShut {NoStop}%
\bibitem [{\citenamefont {{Hu}}(1972)}]{Hu1972}%
  \BibitemOpen
  \bibfield  {author} {\bibinfo {author} {\bibfnamefont {C.-R.}\ \bibnamefont
  {{Hu}}},\ }\href {https://doi.org/10.1103/PhysRevB.6.1756} {\bibfield
  {journal} {\bibinfo  {journal} {\prb}\ }\textbf {\bibinfo {volume} {6}},\
  \bibinfo {pages} {1756} (\bibinfo {year} {1972})}\BibitemShut {NoStop}%
\bibitem [{\citenamefont {{Brandt}}(2003)}]{Brandt2003}%
  \BibitemOpen
  \bibfield  {author} {\bibinfo {author} {\bibfnamefont {E.~H.}\ \bibnamefont
  {{Brandt}}},\ }\href {https://doi.org/10.1103/PhysRevB.68.054506} {\bibfield
  {journal} {\bibinfo  {journal} {\prb}\ }\textbf {\bibinfo {volume} {68}},\
  \bibinfo {eid} {054506} (\bibinfo {year} {2003})},\ \Eprint
  {https://arxiv.org/abs/cond-mat/0304237} {arXiv:cond-mat/0304237
  [cond-mat.supr-con]} \BibitemShut {NoStop}%
\bibitem [{\citenamefont {{Sedrakian}}\ \emph {et~al.}(1997)\citenamefont
  {{Sedrakian}}, \citenamefont {{Sedrakian}},\ and\ \citenamefont
  {{Zharkov}}}]{Sedrakian1997MNRAS}%
  \BibitemOpen
  \bibfield  {author} {\bibinfo {author} {\bibfnamefont {D.~M.}\ \bibnamefont
  {{Sedrakian}}}, \bibinfo {author} {\bibfnamefont {A.~D.}\ \bibnamefont
  {{Sedrakian}}},\ and\ \bibinfo {author} {\bibfnamefont {G.~F.}\ \bibnamefont
  {{Zharkov}}},\ }\href {https://doi.org/10.1093/mnras/290.1.203} {\bibfield
  {journal} {\bibinfo  {journal} {\mnras}\ }\textbf {\bibinfo {volume} {290}},\
  \bibinfo {pages} {203} (\bibinfo {year} {1997})},\ \Eprint
  {https://arxiv.org/abs/astro-ph/9710280} {arXiv:astro-ph/9710280 [astro-ph]}
  \BibitemShut {NoStop}%
\bibitem [{\citenamefont {{Sedrakian}}(2005)}]{Sedrakian2005PhRvD}%
  \BibitemOpen
  \bibfield  {author} {\bibinfo {author} {\bibfnamefont {A.}~\bibnamefont
  {{Sedrakian}}},\ }\href {https://doi.org/10.1103/PhysRevD.71.083003}
  {\bibfield  {journal} {\bibinfo  {journal} {\prd}\ }\textbf {\bibinfo
  {volume} {71}},\ \bibinfo {eid} {083003} (\bibinfo {year} {2005})},\ \Eprint
  {https://arxiv.org/abs/astro-ph/0408467} {arXiv:astro-ph/0408467 [astro-ph]}
  \BibitemShut {NoStop}%
\bibitem [{\citenamefont {{Braithwaite}}(2009)}]{BR2009}%
  \BibitemOpen
  \bibfield  {author} {\bibinfo {author} {\bibfnamefont {J.}~\bibnamefont
  {{Braithwaite}}},\ }\href {https://doi.org/10.1111/j.1365-2966.2008.14034.x}
  {\bibfield  {journal} {\bibinfo  {journal} {\mnras}\ }\textbf {\bibinfo
  {volume} {397}},\ \bibinfo {pages} {763} (\bibinfo {year} {2009})},\ \Eprint
  {https://arxiv.org/abs/0810.1049} {arXiv:0810.1049 [astro-ph]} \BibitemShut
  {NoStop}%
\bibitem [{\citenamefont {{Das}}\ and\ \citenamefont
  {{Mukhopadhyay}}(2023)}]{das-mukhopadhyay}%
  \BibitemOpen
  \bibfield  {author} {\bibinfo {author} {\bibfnamefont {M.}~\bibnamefont
  {{Das}}}\ and\ \bibinfo {author} {\bibfnamefont {B.}~\bibnamefont
  {{Mukhopadhyay}}},\ }\href {https://doi.org/10.3847/1538-4357/aceb63}
  {\bibfield  {journal} {\bibinfo  {journal} {\apj}\ }\textbf {\bibinfo
  {volume} {955}},\ \bibinfo {eid} {19} (\bibinfo {year} {2023})},\ \Eprint
  {https://arxiv.org/abs/2302.03706} {arXiv:2302.03706 [astro-ph.HE]}
  \BibitemShut {NoStop}%
\bibitem [{\citenamefont {{Easson}}\ and\ \citenamefont
  {{Pethick}}(1977)}]{easson1977}%
  \BibitemOpen
  \bibfield  {author} {\bibinfo {author} {\bibfnamefont {I.}~\bibnamefont
  {{Easson}}}\ and\ \bibinfo {author} {\bibfnamefont {C.~J.}\ \bibnamefont
  {{Pethick}}},\ }\href {https://doi.org/10.1103/PhysRevD.16.275} {\bibfield
  {journal} {\bibinfo  {journal} {\prd}\ }\textbf {\bibinfo {volume} {16}},\
  \bibinfo {pages} {275} (\bibinfo {year} {1977})}\BibitemShut {NoStop}%
\bibitem [{\citenamefont {{Frieben}}\ and\ \citenamefont
  {{Rezzolla}}(2012)}]{freibenrezz2012}%
  \BibitemOpen
  \bibfield  {author} {\bibinfo {author} {\bibfnamefont {J.}~\bibnamefont
  {{Frieben}}}\ and\ \bibinfo {author} {\bibfnamefont {L.}~\bibnamefont
  {{Rezzolla}}},\ }\href {https://doi.org/10.1111/j.1365-2966.2012.22027.x}
  {\bibfield  {journal} {\bibinfo  {journal} {\mnras}\ }\textbf {\bibinfo
  {volume} {427}},\ \bibinfo {pages} {3406} (\bibinfo {year} {2012})},\ \Eprint
  {https://arxiv.org/abs/1207.4035} {arXiv:1207.4035 [gr-qc]} \BibitemShut
  {NoStop}%
\bibitem [{\citenamefont {{Soldateschi}}\ and\ \citenamefont
  {{Bucciantini}}(2021)}]{sold2021}%
  \BibitemOpen
  \bibfield  {author} {\bibinfo {author} {\bibfnamefont {J.}~\bibnamefont
  {{Soldateschi}}}\ and\ \bibinfo {author} {\bibfnamefont {N.}~\bibnamefont
  {{Bucciantini}}},\ }\href {https://doi.org/10.3390/galaxies9040101}
  {\bibfield  {journal} {\bibinfo  {journal} {Galaxies}\ }\textbf {\bibinfo
  {volume} {9}},\ \bibinfo {eid} {101} (\bibinfo {year} {2021})},\ \Eprint
  {https://arxiv.org/abs/2110.06039} {arXiv:2110.06039 [astro-ph.HE]}
  \BibitemShut {NoStop}%
\bibitem [{\citenamefont {{Smith}}\ \emph {et~al.}(2023)\citenamefont
  {{Smith}}, \citenamefont {{Abdollahi}}, \citenamefont {{Ajello}},
  \citenamefont {{Bailes}}, \citenamefont {{Baldini}}, \citenamefont
  {{Ballet}}, \citenamefont {{Baring}}, \citenamefont {{Bassa}}, \citenamefont
  {{Gonzalez}}, \citenamefont {{Bellazzini}}, \citenamefont {{Berretta}},
  \citenamefont {{Bhattacharyya}}, \citenamefont {{Bissaldi}}, \citenamefont
  {{Bonino}}, \citenamefont {{Bottacini}}, \citenamefont {{Bregeon}},
  \citenamefont {{Bruel}}, \citenamefont {{Burgay}}, \citenamefont {{Burnett}},
  \citenamefont {{Cameron}}, \citenamefont {{Camilo}}, \citenamefont
  {{Caputo}}, \citenamefont {{Caraveo}}, \citenamefont {{Cavazzuti}},
  \citenamefont {{Chiaro}}, \citenamefont {{Ciprini}}, \citenamefont {{Clark}},
  \citenamefont {{Cognard}}, \citenamefont {{Corongiu}}, \citenamefont
  {{Orestano}}, \citenamefont {{Crnogorcevic}}, \citenamefont {{Cuoco}},
  \citenamefont {{Cutini}}, \citenamefont {{D'Ammando}}, \citenamefont {{de
  Angelis}}, \citenamefont {{DeCesar}}, \citenamefont {{De Gaetano}},
  \citenamefont {{de Menezes}}, \citenamefont {{Deneva}}, \citenamefont {{de
  Palma}}, \citenamefont {{Di Lalla}}, \citenamefont {{Dirirsa}}, \citenamefont
  {{Di Venere}}, \citenamefont {{Dom{\'\i}nguez}}, \citenamefont {{Dumora}},
  \citenamefont {{Fegan}}, \citenamefont {{Ferrara}}, \citenamefont {{Fiori}},
  \citenamefont {{Fleischhack}}, \citenamefont {{Flynn}}, \citenamefont
  {{Franckowiak}}, \citenamefont {{Freire}}, \citenamefont {{Fukazawa}},
  \citenamefont {{Fusco}}, \citenamefont {{Galanti}}, \citenamefont
  {{Gammaldi}}, \citenamefont {{Gargano}}, \citenamefont {{Gasparrini}},
  \citenamefont {{Giacchino}}, \citenamefont {{Giglietto}}, \citenamefont
  {{Giordano}}, \citenamefont {{Giroletti}}, \citenamefont {{Green}},
  \citenamefont {{Grenier}}, \citenamefont {{Guillemot}}, \citenamefont
  {{Guiriec}}, \citenamefont {{Gustafsson}}, \citenamefont {{Harding}},
  \citenamefont {{Hays}}, \citenamefont {{Hewitt}}, \citenamefont {{Horan}},
  \citenamefont {{Hou}}, \citenamefont {{Jankowski}}, \citenamefont
  {{Johnson}}, \citenamefont {{Johnson}}, \citenamefont {{Johnston}},
  \citenamefont {{Kataoka}}, \citenamefont {{Keith}}, \citenamefont {{Kerr}},
  \citenamefont {{Kramer}}, \citenamefont {{Kuss}}, \citenamefont
  {{Latronico}}, \citenamefont {{Lee}}, \citenamefont {{Li}}, \citenamefont
  {{Li}}, \citenamefont {{Limyansky}}, \citenamefont {{Longo}}, \citenamefont
  {{Loparco}}, \citenamefont {{Lorusso}}, \citenamefont {{Lovellette}},
  \citenamefont {{Lower}}, \citenamefont {{Lubrano}}, \citenamefont {{Lyne}},
  \citenamefont {{Maan}}, \citenamefont {{Maldera}}, \citenamefont
  {{Manchester}}, \citenamefont {{Manfreda}}, \citenamefont {{Marelli}},
  \citenamefont {{Mart{\'\i}-Devesa}}, \citenamefont {{Mazziotta}},
  \citenamefont {{McEnery}}, \citenamefont {{Mereu}}, \citenamefont
  {{Michelson}}, \citenamefont {{Mickaliger}}, \citenamefont {{Mitthumsiri}},
  \citenamefont {{Mizuno}}, \citenamefont {{Moiseev}}, \citenamefont
  {{Monzani}}, \citenamefont {{Morselli}}, \citenamefont {{Negro}},
  \citenamefont {{Nemmen}}, \citenamefont {{Nieder}}, \citenamefont {{Nuss}},
  \citenamefont {{Omodei}}, \citenamefont {{Orienti}}, \citenamefont
  {{Orlando}}, \citenamefont {{Ormes}}, \citenamefont {{Palatiello}},
  \citenamefont {{Paneque}}, \citenamefont {{Panzarini}}, \citenamefont
  {{Parthasarathy}}, \citenamefont {{Persic}}, \citenamefont {{Pesce-Rollins}},
  \citenamefont {{Pillera}}, \citenamefont {{Poon}}, \citenamefont {{Porter}},
  \citenamefont {{Possenti}}, \citenamefont {{Principe}}, \citenamefont
  {{Rain{\`o}}}, \citenamefont {{Rando}}, \citenamefont {{Ransom}},
  \citenamefont {{Ray}}, \citenamefont {{Razzano}}, \citenamefont {{Razzaque}},
  \citenamefont {{Reimer}}, \citenamefont {{Reimer}}, \citenamefont
  {{Renault-Tinacci}}, \citenamefont {{Romani}}, \citenamefont
  {{S{\'a}nchez-Conde}}, \citenamefont {{Parkinson}}, \citenamefont
  {{Scotton}}, \citenamefont {{Serini}}, \citenamefont {{Sgr{\`o}}},
  \citenamefont {{Shannon}}, \citenamefont {{Sharma}}, \citenamefont {{Shen}},
  \citenamefont {{Siskind}}, \citenamefont {{Spandre}}, \citenamefont
  {{Spinelli}}, \citenamefont {{Stappers}}, \citenamefont {{Stephens}},
  \citenamefont {{Suson}}, \citenamefont {{Tabassum}}, \citenamefont
  {{Tajima}}, \citenamefont {{Tak}}, \citenamefont {{Theureau}}, \citenamefont
  {{Thompson}}, \citenamefont {{Tibolla}}, \citenamefont {{Torres}},
  \citenamefont {{Valverde}}, \citenamefont {{Venter}}, \citenamefont
  {{Wadiasingh}}, \citenamefont {{Wang}}, \citenamefont {{Wang}}, \citenamefont
  {{Wang}}, \citenamefont {{Weltevrede}}, \citenamefont {{Wood}}, \citenamefont
  {{Yan}}, \citenamefont {{Zaharijas}}, \citenamefont {{Zhang}},\ and\
  \citenamefont {{Zhu}}}]{fermilat}%
  \BibitemOpen
  \bibfield  {author} {\bibinfo {author} {\bibfnamefont {D.~A.}\ \bibnamefont
  {{Smith}}}, \bibinfo {author} {\bibfnamefont {S.}~\bibnamefont
  {{Abdollahi}}}, \bibinfo {author} {\bibfnamefont {M.}~\bibnamefont
  {{Ajello}}}, \bibinfo {author} {\bibfnamefont {M.}~\bibnamefont {{Bailes}}},
  \bibinfo {author} {\bibfnamefont {L.}~\bibnamefont {{Baldini}}}, \bibinfo
  {author} {\bibfnamefont {J.}~\bibnamefont {{Ballet}}}, \bibinfo {author}
  {\bibfnamefont {M.~G.}\ \bibnamefont {{Baring}}}, \bibinfo {author}
  {\bibfnamefont {C.}~\bibnamefont {{Bassa}}}, \bibinfo {author} {\bibfnamefont
  {J.~B.}\ \bibnamefont {{Gonzalez}}}, \bibinfo {author} {\bibfnamefont
  {R.}~\bibnamefont {{Bellazzini}}}, \bibinfo {author} {\bibfnamefont
  {A.}~\bibnamefont {{Berretta}}}, \bibinfo {author} {\bibfnamefont
  {B.}~\bibnamefont {{Bhattacharyya}}}, \bibinfo {author} {\bibfnamefont
  {E.}~\bibnamefont {{Bissaldi}}}, \bibinfo {author} {\bibfnamefont
  {R.}~\bibnamefont {{Bonino}}}, \bibinfo {author} {\bibfnamefont
  {E.}~\bibnamefont {{Bottacini}}}, \bibinfo {author} {\bibfnamefont
  {J.}~\bibnamefont {{Bregeon}}}, \bibinfo {author} {\bibfnamefont
  {P.}~\bibnamefont {{Bruel}}}, \bibinfo {author} {\bibfnamefont
  {M.}~\bibnamefont {{Burgay}}}, \bibinfo {author} {\bibfnamefont {T.~H.}\
  \bibnamefont {{Burnett}}}, \bibinfo {author} {\bibfnamefont {R.~A.}\
  \bibnamefont {{Cameron}}}, \bibinfo {author} {\bibfnamefont {F.}~\bibnamefont
  {{Camilo}}}, \bibinfo {author} {\bibfnamefont {R.}~\bibnamefont {{Caputo}}},
  \bibinfo {author} {\bibfnamefont {P.~A.}\ \bibnamefont {{Caraveo}}}, \bibinfo
  {author} {\bibfnamefont {E.}~\bibnamefont {{Cavazzuti}}}, \bibinfo {author}
  {\bibfnamefont {G.}~\bibnamefont {{Chiaro}}}, \bibinfo {author}
  {\bibfnamefont {S.}~\bibnamefont {{Ciprini}}}, \bibinfo {author}
  {\bibfnamefont {C.~J.}\ \bibnamefont {{Clark}}}, \bibinfo {author}
  {\bibfnamefont {I.}~\bibnamefont {{Cognard}}}, \bibinfo {author}
  {\bibfnamefont {A.}~\bibnamefont {{Corongiu}}}, \bibinfo {author}
  {\bibfnamefont {P.~C.}\ \bibnamefont {{Orestano}}}, \bibinfo {author}
  {\bibfnamefont {M.}~\bibnamefont {{Crnogorcevic}}}, \bibinfo {author}
  {\bibfnamefont {A.}~\bibnamefont {{Cuoco}}}, \bibinfo {author} {\bibfnamefont
  {S.}~\bibnamefont {{Cutini}}}, \bibinfo {author} {\bibfnamefont
  {F.}~\bibnamefont {{D'Ammando}}}, \bibinfo {author} {\bibfnamefont
  {A.}~\bibnamefont {{de Angelis}}}, \bibinfo {author} {\bibfnamefont {M.~E.}\
  \bibnamefont {{DeCesar}}}, \bibinfo {author} {\bibfnamefont {S.}~\bibnamefont
  {{De Gaetano}}}, \bibinfo {author} {\bibfnamefont {R.}~\bibnamefont {{de
  Menezes}}}, \bibinfo {author} {\bibfnamefont {J.}~\bibnamefont {{Deneva}}},
  \bibinfo {author} {\bibfnamefont {F.}~\bibnamefont {{de Palma}}}, \bibinfo
  {author} {\bibfnamefont {N.}~\bibnamefont {{Di Lalla}}}, \bibinfo {author}
  {\bibfnamefont {F.}~\bibnamefont {{Dirirsa}}}, \bibinfo {author}
  {\bibfnamefont {L.}~\bibnamefont {{Di Venere}}}, \bibinfo {author}
  {\bibfnamefont {A.}~\bibnamefont {{Dom{\'\i}nguez}}}, \bibinfo {author}
  {\bibfnamefont {D.}~\bibnamefont {{Dumora}}}, \bibinfo {author}
  {\bibfnamefont {S.~J.}\ \bibnamefont {{Fegan}}}, \bibinfo {author}
  {\bibfnamefont {E.~C.}\ \bibnamefont {{Ferrara}}}, \bibinfo {author}
  {\bibfnamefont {A.}~\bibnamefont {{Fiori}}}, \bibinfo {author} {\bibfnamefont
  {H.}~\bibnamefont {{Fleischhack}}}, \bibinfo {author} {\bibfnamefont
  {C.}~\bibnamefont {{Flynn}}}, \bibinfo {author} {\bibfnamefont
  {A.}~\bibnamefont {{Franckowiak}}}, \bibinfo {author} {\bibfnamefont
  {P.~C.~C.}\ \bibnamefont {{Freire}}}, \bibinfo {author} {\bibfnamefont
  {Y.}~\bibnamefont {{Fukazawa}}}, \bibinfo {author} {\bibfnamefont
  {P.}~\bibnamefont {{Fusco}}}, \bibinfo {author} {\bibfnamefont
  {G.}~\bibnamefont {{Galanti}}}, \bibinfo {author} {\bibfnamefont
  {V.}~\bibnamefont {{Gammaldi}}}, \bibinfo {author} {\bibfnamefont
  {F.}~\bibnamefont {{Gargano}}}, \bibinfo {author} {\bibfnamefont
  {D.}~\bibnamefont {{Gasparrini}}}, \bibinfo {author} {\bibfnamefont
  {F.}~\bibnamefont {{Giacchino}}}, \bibinfo {author} {\bibfnamefont
  {N.}~\bibnamefont {{Giglietto}}}, \bibinfo {author} {\bibfnamefont
  {F.}~\bibnamefont {{Giordano}}}, \bibinfo {author} {\bibfnamefont
  {M.}~\bibnamefont {{Giroletti}}}, \bibinfo {author} {\bibfnamefont
  {D.}~\bibnamefont {{Green}}}, \bibinfo {author} {\bibfnamefont {I.~A.}\
  \bibnamefont {{Grenier}}}, \bibinfo {author} {\bibfnamefont {L.}~\bibnamefont
  {{Guillemot}}}, \bibinfo {author} {\bibfnamefont {S.}~\bibnamefont
  {{Guiriec}}}, \bibinfo {author} {\bibfnamefont {M.}~\bibnamefont
  {{Gustafsson}}}, \bibinfo {author} {\bibfnamefont {A.~K.}\ \bibnamefont
  {{Harding}}}, \bibinfo {author} {\bibfnamefont {E.}~\bibnamefont {{Hays}}},
  \bibinfo {author} {\bibfnamefont {J.~W.}\ \bibnamefont {{Hewitt}}}, \bibinfo
  {author} {\bibfnamefont {D.}~\bibnamefont {{Horan}}}, \bibinfo {author}
  {\bibfnamefont {X.}~\bibnamefont {{Hou}}}, \bibinfo {author} {\bibfnamefont
  {F.}~\bibnamefont {{Jankowski}}}, \bibinfo {author} {\bibfnamefont {R.~P.}\
  \bibnamefont {{Johnson}}}, \bibinfo {author} {\bibfnamefont {T.~J.}\
  \bibnamefont {{Johnson}}}, \bibinfo {author} {\bibfnamefont {S.}~\bibnamefont
  {{Johnston}}}, \bibinfo {author} {\bibfnamefont {J.}~\bibnamefont
  {{Kataoka}}}, \bibinfo {author} {\bibfnamefont {M.~J.}\ \bibnamefont
  {{Keith}}}, \bibinfo {author} {\bibfnamefont {M.}~\bibnamefont {{Kerr}}},
  \bibinfo {author} {\bibfnamefont {M.}~\bibnamefont {{Kramer}}}, \bibinfo
  {author} {\bibfnamefont {M.}~\bibnamefont {{Kuss}}}, \bibinfo {author}
  {\bibfnamefont {L.}~\bibnamefont {{Latronico}}}, \bibinfo {author}
  {\bibfnamefont {S.~H.}\ \bibnamefont {{Lee}}}, \bibinfo {author}
  {\bibfnamefont {D.}~\bibnamefont {{Li}}}, \bibinfo {author} {\bibfnamefont
  {J.}~\bibnamefont {{Li}}}, \bibinfo {author} {\bibfnamefont {B.}~\bibnamefont
  {{Limyansky}}}, \bibinfo {author} {\bibfnamefont {F.}~\bibnamefont
  {{Longo}}}, \bibinfo {author} {\bibfnamefont {F.}~\bibnamefont {{Loparco}}},
  \bibinfo {author} {\bibfnamefont {L.}~\bibnamefont {{Lorusso}}}, \bibinfo
  {author} {\bibfnamefont {M.~N.}\ \bibnamefont {{Lovellette}}}, \bibinfo
  {author} {\bibfnamefont {M.}~\bibnamefont {{Lower}}}, \bibinfo {author}
  {\bibfnamefont {P.}~\bibnamefont {{Lubrano}}}, \bibinfo {author}
  {\bibfnamefont {A.~G.}\ \bibnamefont {{Lyne}}}, \bibinfo {author}
  {\bibfnamefont {Y.}~\bibnamefont {{Maan}}}, \bibinfo {author} {\bibfnamefont
  {S.}~\bibnamefont {{Maldera}}}, \bibinfo {author} {\bibfnamefont {R.~N.}\
  \bibnamefont {{Manchester}}}, \bibinfo {author} {\bibfnamefont
  {A.}~\bibnamefont {{Manfreda}}}, \bibinfo {author} {\bibfnamefont
  {M.}~\bibnamefont {{Marelli}}}, \bibinfo {author} {\bibfnamefont
  {G.}~\bibnamefont {{Mart{\'\i}-Devesa}}}, \bibinfo {author} {\bibfnamefont
  {M.~N.}\ \bibnamefont {{Mazziotta}}}, \bibinfo {author} {\bibfnamefont
  {J.~E.}\ \bibnamefont {{McEnery}}}, \bibinfo {author} {\bibfnamefont
  {I.}~\bibnamefont {{Mereu}}}, \bibinfo {author} {\bibfnamefont {P.~F.}\
  \bibnamefont {{Michelson}}}, \bibinfo {author} {\bibfnamefont
  {M.}~\bibnamefont {{Mickaliger}}}, \bibinfo {author} {\bibfnamefont
  {W.}~\bibnamefont {{Mitthumsiri}}}, \bibinfo {author} {\bibfnamefont
  {T.}~\bibnamefont {{Mizuno}}}, \bibinfo {author} {\bibfnamefont {A.~A.}\
  \bibnamefont {{Moiseev}}}, \bibinfo {author} {\bibfnamefont {M.~E.}\
  \bibnamefont {{Monzani}}}, \bibinfo {author} {\bibfnamefont {A.}~\bibnamefont
  {{Morselli}}}, \bibinfo {author} {\bibfnamefont {M.}~\bibnamefont {{Negro}}},
  \bibinfo {author} {\bibfnamefont {R.}~\bibnamefont {{Nemmen}}}, \bibinfo
  {author} {\bibfnamefont {L.}~\bibnamefont {{Nieder}}}, \bibinfo {author}
  {\bibfnamefont {E.}~\bibnamefont {{Nuss}}}, \bibinfo {author} {\bibfnamefont
  {N.}~\bibnamefont {{Omodei}}}, \bibinfo {author} {\bibfnamefont
  {M.}~\bibnamefont {{Orienti}}}, \bibinfo {author} {\bibfnamefont
  {E.}~\bibnamefont {{Orlando}}}, \bibinfo {author} {\bibfnamefont {J.~F.}\
  \bibnamefont {{Ormes}}}, \bibinfo {author} {\bibfnamefont {M.}~\bibnamefont
  {{Palatiello}}}, \bibinfo {author} {\bibfnamefont {D.}~\bibnamefont
  {{Paneque}}}, \bibinfo {author} {\bibfnamefont {G.}~\bibnamefont
  {{Panzarini}}}, \bibinfo {author} {\bibfnamefont {A.}~\bibnamefont
  {{Parthasarathy}}}, \bibinfo {author} {\bibfnamefont {M.}~\bibnamefont
  {{Persic}}}, \bibinfo {author} {\bibfnamefont {M.}~\bibnamefont
  {{Pesce-Rollins}}}, \bibinfo {author} {\bibfnamefont {R.}~\bibnamefont
  {{Pillera}}}, \bibinfo {author} {\bibfnamefont {H.}~\bibnamefont {{Poon}}},
  \bibinfo {author} {\bibfnamefont {T.~A.}\ \bibnamefont {{Porter}}}, \bibinfo
  {author} {\bibfnamefont {A.}~\bibnamefont {{Possenti}}}, \bibinfo {author}
  {\bibfnamefont {G.}~\bibnamefont {{Principe}}}, \bibinfo {author}
  {\bibfnamefont {S.}~\bibnamefont {{Rain{\`o}}}}, \bibinfo {author}
  {\bibfnamefont {R.}~\bibnamefont {{Rando}}}, \bibinfo {author} {\bibfnamefont
  {S.~M.}\ \bibnamefont {{Ransom}}}, \bibinfo {author} {\bibfnamefont {P.~S.}\
  \bibnamefont {{Ray}}}, \bibinfo {author} {\bibfnamefont {M.}~\bibnamefont
  {{Razzano}}}, \bibinfo {author} {\bibfnamefont {S.}~\bibnamefont
  {{Razzaque}}}, \bibinfo {author} {\bibfnamefont {A.}~\bibnamefont
  {{Reimer}}}, \bibinfo {author} {\bibfnamefont {O.}~\bibnamefont {{Reimer}}},
  \bibinfo {author} {\bibfnamefont {N.}~\bibnamefont {{Renault-Tinacci}}},
  \bibinfo {author} {\bibfnamefont {R.~W.}\ \bibnamefont {{Romani}}}, \bibinfo
  {author} {\bibfnamefont {M.}~\bibnamefont {{S{\'a}nchez-Conde}}}, \bibinfo
  {author} {\bibfnamefont {P.~M.~S.}\ \bibnamefont {{Parkinson}}}, \bibinfo
  {author} {\bibfnamefont {L.}~\bibnamefont {{Scotton}}}, \bibinfo {author}
  {\bibfnamefont {D.}~\bibnamefont {{Serini}}}, \bibinfo {author}
  {\bibfnamefont {C.}~\bibnamefont {{Sgr{\`o}}}}, \bibinfo {author}
  {\bibfnamefont {R.}~\bibnamefont {{Shannon}}}, \bibinfo {author}
  {\bibfnamefont {V.}~\bibnamefont {{Sharma}}}, \bibinfo {author}
  {\bibfnamefont {Z.}~\bibnamefont {{Shen}}}, \bibinfo {author} {\bibfnamefont
  {E.~J.}\ \bibnamefont {{Siskind}}}, \bibinfo {author} {\bibfnamefont
  {G.}~\bibnamefont {{Spandre}}}, \bibinfo {author} {\bibfnamefont
  {P.}~\bibnamefont {{Spinelli}}}, \bibinfo {author} {\bibfnamefont {B.~W.}\
  \bibnamefont {{Stappers}}}, \bibinfo {author} {\bibfnamefont {T.~E.}\
  \bibnamefont {{Stephens}}}, \bibinfo {author} {\bibfnamefont {D.~J.}\
  \bibnamefont {{Suson}}}, \bibinfo {author} {\bibfnamefont {S.}~\bibnamefont
  {{Tabassum}}}, \bibinfo {author} {\bibfnamefont {H.}~\bibnamefont
  {{Tajima}}}, \bibinfo {author} {\bibfnamefont {D.}~\bibnamefont {{Tak}}},
  \bibinfo {author} {\bibfnamefont {G.}~\bibnamefont {{Theureau}}}, \bibinfo
  {author} {\bibfnamefont {D.~J.}\ \bibnamefont {{Thompson}}}, \bibinfo
  {author} {\bibfnamefont {O.}~\bibnamefont {{Tibolla}}}, \bibinfo {author}
  {\bibfnamefont {D.~F.}\ \bibnamefont {{Torres}}}, \bibinfo {author}
  {\bibfnamefont {J.}~\bibnamefont {{Valverde}}}, \bibinfo {author}
  {\bibfnamefont {C.}~\bibnamefont {{Venter}}}, \bibinfo {author}
  {\bibfnamefont {Z.}~\bibnamefont {{Wadiasingh}}}, \bibinfo {author}
  {\bibfnamefont {N.}~\bibnamefont {{Wang}}}, \bibinfo {author} {\bibfnamefont
  {N.}~\bibnamefont {{Wang}}}, \bibinfo {author} {\bibfnamefont
  {P.}~\bibnamefont {{Wang}}}, \bibinfo {author} {\bibfnamefont
  {P.}~\bibnamefont {{Weltevrede}}}, \bibinfo {author} {\bibfnamefont
  {K.}~\bibnamefont {{Wood}}}, \bibinfo {author} {\bibfnamefont
  {J.}~\bibnamefont {{Yan}}}, \bibinfo {author} {\bibfnamefont
  {G.}~\bibnamefont {{Zaharijas}}}, \bibinfo {author} {\bibfnamefont
  {C.}~\bibnamefont {{Zhang}}},\ and\ \bibinfo {author} {\bibfnamefont
  {W.}~\bibnamefont {{Zhu}}},\ }\href
  {https://doi.org/10.3847/1538-4357/acee67} {\bibfield  {journal} {\bibinfo
  {journal} {\apj}\ }\textbf {\bibinfo {volume} {958}},\ \bibinfo {eid} {191}
  (\bibinfo {year} {2023})},\ \Eprint {https://arxiv.org/abs/2307.11132}
  {arXiv:2307.11132 [astro-ph.HE]} \BibitemShut {NoStop}%
\end{thebibliography}%

\end{document}